\documentclass[10pt]{iopart}

\usepackage{graphicx}

\def\st#1{{\kern-4pt} \not\!#1}
\def\sp{\kern +3pt}
\def\sm{\kern -3pt}

\def\spQ{\kern +6pt}

\def\bea{\begin{eqnarray}}
\def\eea{\end{eqnarray}}

\def\sfrac#1#2{{\textstyle \frac{#1}{#2}}}

\newcommand{\bra}[1]{\langle #1|}
\newcommand{\ket}[1]{|#1\rangle}

\def\be{\begin{equation}}
\def\ee{\end{equation}}
\def\ba{\begin{eqnarray}}
\def\ea{\end{eqnarray}}

\begin{document}

\title[Octet baryon electromagnetic form factors in nuclear medium
]{Octet baryon electromagnetic form factors in nuclear medium}

\author{G. Ramalho$^1$, K. Tsushima$^2$,  and A.~W.~Thomas$^{2,3}$}

\address{$^1$
CFTP, Instituto Superior T\'ecnico,
Universidade T\'ecnica de Lisboa,
Av.~Rovisco Pais, 1049-001 Lisboa, Portugal
}
\address{
$^2$CSSM and $^3$CoEPP, School of Chemistry and Physics,
University of Adelaide, Adelaide SA 5005, Australia
}
\begin{abstract}
We study the octet baryon electromagnetic form factors
in nuclear matter
using the covariant spectator quark model extended
to the nuclear matter regime.
The parameters of the model in vacuum are fixed
by the study of the octet baryon electromagnetic form factors.
In nuclear matter the changes in hadron properties
are calculated by including the relevant hadron masses and the
modification of the pion-baryon coupling constants
calculated in the quark-meson coupling model.
In nuclear matter the magnetic form factors of the octet baryons
are enhanced in the low $Q^2$ region, while the electric
form factors show a more rapid variation with $Q^2$.
The results are compared with the modification of the
bound proton electromagnetic form factors observed at Jefferson Lab.
In addition, the corresponding changes for the bound neutron are predicted.
\end{abstract}

\pacs{13.40.Gp, 21.65.-f, 14.20.Jn, 12.39.Ki}

\eqnobysec

\section{Introduction}

Whether or not hadrons change their properties in
a nuclear medium, has been one of the long-standing problems in nuclear
physics~\cite{BrownRho,JLabbook}. QCD is established as the theory
of strong interactions and quarks and gluons are the
degrees of freedom in the QCD Lagrangian.
It seems natural that in the strong mean fields which
pervade nuclear matter the motion of the quarks and gluons inside
hadrons should be modified. Such changes are what is meant by
the nuclear modification of hadron properties and studying such
effect is clearly central to the understanding of
dense matter within QCD.

Recently, strong evidence concerning the modification of
nucleon properties in a nuclear medium has been reported
from the proton electromagnetic form factors measured
in polarized ($\vec{e},e'\vec{p}$) scattering on $^{16}$O~\cite{eepO16}
and $^4$He~\cite{Dieterich01,Strauch03,Paolone10,Malace11}
at MAMI and Jefferson Lab.
These experiments measured the double ratio of proton-recoil
polarization transfer coefficients in the quasi-elastic scattering off nuclei,
and the results were normalized
with respective to the double ratio of hydrogen.
The results from $^4$He strongly suggest the modification of
the bound proton electromagnetic form factors.
Furthermore, the study of neutron properties in the nuclear medium
in  \cite{Cloet09}, predicts an enhancement of the
same double ratio for the neutron, contrary to the suppression observed
for the proton. The corresponding experiment, to measure the
polarization transfer, is planned in the future~\cite{npolarization}.
Theoretically, there are several studies concerning
the electromagnetic form factors of the nucleon
in the nuclear medium~\cite{Cloet09,QMCEMFFMedium,Miller,Bentz,MEC}.
They are based either on quark degrees of
freedom~\cite{Cloet09,QMCEMFFMedium,Miller,Bentz},
or meson and nucleon degrees of freedom~\cite{MEC}.
However, it is very difficult to separate and identify
the observed effects in terms of these degrees of freedom.
In particular, to distinguish a possible change in the nucleon
properties in a nuclear medium from those of the conventional
many-body effects, such as final state interactions and meson exchange current,
are very difficult, and seems only possible in a model dependent way,
where experimental measurement involves all such effects
including the one-body current modification~\cite{JLabbook,Tsushima93,Tsushima03}.
Thus, the interpretation of the modification observed
is still under discussion and has not been settled yet.
In these circumstances it is helpful to examine the modification
of the electromagnetic form factors of the nucleon and other baryons
in the nuclear medium within alternative approaches.

In this article we study the medium modification of the
the octet baryon electromagnetic form factors in nuclear matter
focusing on the valence quark structure of a baryon.
Thus, we do not include final state interactions nor meson exchange current,
where the latter may possibly be important for the magnetic form factors
of baryons in a nucleus and nuclear matter~\cite{MEC,Tsushima93,MEC1,MEC2,MEC3,MEC4}.
For this purpose, we use the
covariant spectator
quark model~\cite{Nucleon,Nucleon2,NucleonDIS,FixedAxis,NDelta},
which has its basis in the covariant spectator theory~\cite{Gross}.
The model has been successfully applied to study the electromagnetic
properties of the octet~\cite{OctetMag,OctetEMFF} and
decuplet~\cite{DeltaFF,DeltaDFF,DeltaDFF2,Omega,GE2Omega}
baryons. The model was also very successful in the studies of
$\gamma^* N \to \Delta(1232)$~\cite{NDeltaD,LatticeD,Lattice,DeltaTL},
$\gamma^* N \to N(1440)$~\cite{Roper},
and $\gamma^* N \to N(1535)$~\cite{S11} reactions.
For the meson cloud effects, we include the pion cloud effects,
which are expected to be dominant, in a phenomenological manner
based on the method applied in  \cite{NDelta,OctetMag,OctetEMFF}.

In  \cite{OctetEMFF} the model was extended to the lattice regime
to utilize the lattice QCD simulation data,
where the electromagnetic form factors of the octet baryons
were able to be calculated at lattice hadron masses
corresponding to large pion mass values used in
the lattice QCD simulations.
Similarly, it is also possible to extend the model to the
in-medium regime, once we are able to calculate the in-medium
modified masses of the baryons and mesons appearing in the model.
This is the working hypothesis used
to extend the model, and to calculate the in-medium modifications
of the octet baryon electromagnetic form factors.
For the in-medium masses of the baryons and mesons,
we use the quark-meson coupling (QMC) model~\cite{QMC,QMCReview},
which has been successfully applied to study the properties of
nuclei~\cite{QMCNucleiI,QMCNucleiII},
hypernuclei~\cite{QMCHypI,QMCHypII,TsushimaLamb,ShyamHyp},
and hadron properties in a nuclear medium~\cite{Tsushima02b},
based on the relativistic valence quark structure of hadrons
in a nuclear medium.

Another point to note concerning this study is that
the parametrization of the pion cloud contributions in vacuum
has been improved over what was used in the past~\cite{OctetEMFF}
based on lattice QCD simulation and chiral
perturbation theory.
In particular, care is taken for the neutron charge
form factor and the charge radius, where the
pion cloud contributions are very important.
This is also true for the electric charge neutral particles.

This paper is organized as follows.
We start by defining the electromagnetic
form factors in medium in section \ref{secFormFactors}.
In section \ref{secSpectator} we explain the covariant spectator
quark model, and describe the electromagnetic currents
for octet baryons in the model.
The extension of the model to the in-medium
regime is discussed in section \ref{secInMediumRegime}.
Results are presented in section \ref{secResults}, and
discussions and summary are given in section \ref{secDiscussions}.

\section{Electromagnetic form factors in
vacuum and in the nuclear matter}
\label{secFormFactors}

A spin 1/2 baryon $B$, a member of the
baryon octet, has a Dirac structure and
therefore its electromagnetic
structure can be expressed
in terms of two independent
form factors, namely,
the electric $G_{EB}$ and magnetic
$G_{MB}$ form factors in vacuum (mass $M_B$)
and those in the nuclear medium (mass $M_B^\ast$)
$G_{EB}^\ast$ and
$G_{MB}^\ast$, respectively,
and they are defined below.

\subsection{In vacuum}

Let us consider an octet baryon $B$, with mass $M_B$ in vacuum.
When the initial (momentum $P_-$) and the final (momentum $P_+$)
states are on-shell, the electromagnetic current
(coupling of the baryon with a photon) can
be represented as
\be
J_B^\mu=
F_{1B}(Q^2) \gamma^\mu +
F_{2B}(Q^2)
\frac{i \sigma^{\mu \nu}q_\nu}{2M_B},
\label{eqJ0}
\ee
where $q=P_+ -P_-$,
and $F_{1B}$ and $F_{2B}$ are respectively the Dirac and
Pauli form factors which are the functions of $Q^2=-q^2$.

Suppressed in equation (\ref{eqJ0}) are the initial ($u_B$)
and final state ($\bar u_B$) Dirac spinors, functions of $P_\pm$
and the spin projections.
For simplicity we represent the current
in units $e=\sqrt{4\pi\alpha}$, with
$\alpha \simeq 1/137$, the electromagnetic
fine structure constant.

At $Q^2=0$ they are normalized as
\be
F_{1B}(0)= e_B, \hspace{.9cm}
F_{2B}(0)=\kappa_B,
\label{eqCharge0}
\ee
where $e_B$ is the baryon charge in units of $e$
and $\kappa_B$ is the baryon anomalous magnetic moment
in natural units $\sfrac{e}{2M_B}$.

An alternative representation of the
electromagnetic form factors of the baryon $B$
is the Sachs parametrization in terms of the
electric charge $G_E$ and magnetic dipole $G_M$
form factors.
For the electric charge form factor
the following relation holds with $F_{1B}$ and $F_{2B}$,
\ba
G_{EB}(Q^2)= F_{1B}(Q^2) -\frac{Q^2}{4M_B^2}F_{2B}(Q^2).
\label{eqGEB0}
\ea
As for the magnetic dipole form factor $G_{MB}$
the natural definition is $G_{MB}= F_{1B}+ F_{2B}$.
At $Q^2=0$, $G_{MB}(Q^2)$ defines
the magnetic moment of the baryon $B$ in
natural units ($\sfrac{e}{2 M_B}$),
$\mu_B= G_{MB}(0) \sfrac{e}{2 M_B}$.
To compare magnetic moments $\mu_B$
of particles with different masses it is usual
to express $\mu_B$ in terms $\hat \mu_N= \sfrac{e}{2M_N}$,
the nuclear magneton.
In this case  $\mu_B= G_{MB}(0) \frac{M_N}{M_B} \hat \mu_N$.
Therefore, although the study of the
baryon magnetic form factor can be done
naturally using  $G_{MB}= F_{1B}+ F_{2B}$,
as performed in a previous work~\cite{OctetEMFF},
that $\mu_B$ was defined in the natural units,
it is more convenient to define $G_{MB}$
in units of the nuclear magneton.
Throughout this article we will use then
\ba
G_{MB}(Q^2)=
\left[
F_{1B}(Q^2) + F_{2B}(Q^2)\right] \frac{M_N}{M_B}.
\label{eqGMB0}
\ea

\subsection{In medium}

We consider now the octet baryon $B$
in the nuclear medium with the effective mass $M_B^\ast$.
Assuming that the baryon
is quasi-free in the nuclear medium, the
electromagnetic current for
the baryon $B$ can be expressed as
\be
J_B^\mu=
F_{1B}^\ast(Q^2) \gamma^\mu +
F_{2B}^\ast(Q^2)
\frac{i \sigma^{\mu \nu}q_\nu}{2M_B^\ast},
\label{eqJgen}
\ee
where
$F_{1B}^\ast$ and $F_{2B}^\ast$ are respectively the Dirac and
Pauli form factors in the nuclear medium.
Again the in-medium spinors $\bar{u}_B^\ast(P_+)$ and
$u_B^\ast(P_-)$ are suppressed.
At $Q^2=0$, one has also
\be
F_{1B}^\ast(0)= e_B^\ast, \hspace{.9cm}
F_{2B}^\ast(0)=\kappa_B^\ast,
\label{eqCharge}
\ee
where $e_B^\ast$ is the electric charge
in nuclear medium (the same as in the vacuum: $e_B=e_B^\ast$)
and $\kappa_B^\ast$ is the anomalous magnetic moment
in units of $\sfrac{e}{2 M_B^\ast}$.

As in the vacuum [see equations (\ref{eqGEB0})-(\ref{eqGMB0})],
we define the electric charge and magnetic dipole
form factors as
\ba
& &
G_{EB}^\ast(Q^2)= F_{1B}^\ast(Q^2) -\frac{Q^2}{4(M_B^\ast)^2}F_{2B}^\ast(Q^2),
\label{eqGEm}
\\
& &
G_{MB}^\ast(Q^2)=
\left[
F_{1B}^\ast(Q^2) + F_{2B}^\ast(Q^2)\right] \frac{M_N}{M_B^\ast}.
\label{eqGMm}
\ea
Note that the nucleon mass in vacuum ($M_N$)
is included in the definition of $G_{MB}^\ast$.
As mentioned already we use this definition
to make comparison easier with
respect to the vacuum results.

Because the effective nucleon mass is expected to be smaller
than the mass in vacuum, $M_N^\ast < M_N$,
$G_{MN}^\ast(Q^2)$ is expected to
increase and the
magnetic moment is enhanced
in magnitude (\mbox{$|\mu_N^\ast| > |\mu_N|$}).

\section{Spectator quark model}
\label{secSpectator}

We describe now the octet baryon electromagnetic
form factors in vacuum for a baryon $B$ with mass $M_B$
following  \cite{OctetEMFF}.
In next section we describe
the extension of the model to the nuclear medium.

\begin{figure*}[t]
\centerline{
\mbox{
\includegraphics[width=4.0in]{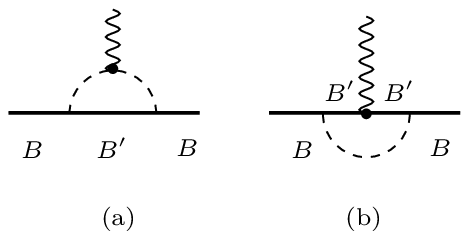}} }
\caption{\footnotesize
Electromagnetic interaction with the baryon $B$
within the one-pion loop level through
the intermediate baryon states $B^\prime$.
A diagram including a contact vertex
$\gamma \pi B B^\prime$, as described in  \cite{OctetMag},
is not represented explicitly, since
the isospin structure is the same as diagram (a).
See  \cite{OctetMag} for details.
}
\label{figPionCloud}
\end{figure*}

The electromagnetic interaction with a baryon $B$ may
be decomposed into the photon interaction
with valence quarks, and with sea quarks
(polarized quark-antiquark pairs or meson cloud).
As the pion is the lightest meson
the pion cloud is expected to give the most important contribution.
Then, one can describe the
electromagnetic interaction for a member $B$
of the octet baryons using a current,
\be
J_B^\mu =
Z_B \left[
J_{0B}^\mu + J_\pi^\mu + J_{\gamma B}^\mu \right],
\label{eqJdecomp}
\ee
where $J_{0B}^\mu$ stands for the electromagnetic
interaction with the quark core without the pion cloud,
and the remaining terms are the interaction
with the intermediate pion-baryon ($\pi B$) states,
as depicted in figure~\ref{figPionCloud}.
In particular, $J_\pi^\mu$ represents the direct interaction
with the pion [diagram (a)], and $J_{\gamma B}^\mu$ the interaction
with the baryon while one pion is in the air [diagram (b)].
The factor $Z_B$  is a renormalization constant,
which is common to each isomultiplet:
nucleon ($N$), $\Sigma$, $\Lambda$, and $\Xi$.
$Z_B$ is related with the derivative of the
baryon self-energy~\cite{OctetMag}.

We restrict the meson cloud dressing to the
pion cloud, since the lightest meson is dominant
as known from chiral perturbation theory.
This is consistent with the studies
of the octet baryon systems~\cite{Wang09,Boinepalli06,Kubodera85,Leinweber04,Franklin02,Kubis99,Cheedket04,Jenkins93,Meissner97,Puglia00,Geng08}.
We note however, that kaon ($K$) cloud contributions
may became more pronounced for systems with
more strangeness, particularly when the pion cloud
contributions are small.

In the previous work~\cite{OctetEMFF} we
presented a model for the valence quark and
meson cloud contributions that
were calibrated by
lattice QCD and physical nucleon
electromagnetic form factor data,
as well as the octet magnetic moment data
($\Lambda$, $\Sigma^{+,-}$ and $\Xi^{0,-}$).
The model provided a good global description
of the octet baryon electromagnetic form factor
data (physical and lattice regimes)
except for the neutron electric form
factor in the low $Q^2$ region.
As a consequence the model underestimated
significantly the neutron electric charge square radius
[$-0.029$ fm$^2$ to be compared with the
experimental result $-0.116$ fm$^2$].
This was interpreted as an insufficient
impact of the nucleon data in the low $Q^2$ region,
particularly those for the neutron.
Another limitation of the model
was no explicit inclusion of the
pion mass ($m_\pi$) in the parametrization of
the pion cloud dressing.
Although the long-range falloff
of the pion cloud effects
in the position space with the distance $r$ going like
$\exp(-2 m_\pi r)$, it can be simulated by
multipole functions with appropriated cutoffs.
In order to study the chiral behavior,
it is vital to include
the pion mass dependence explicitly.

Therefore, we improve the model by adding
two new features
to the model of  \cite{OctetEMFF}:
\begin{itemize}
\item
We constrain the model parameters
using also the experimental values of the
proton and neutron electric and magnetic square radii,
as well as $\Sigma^-$ electric square radius.
\item
We redefine the pion cloud parametrization
in order to reproduce the leading order
chiral behavior for the form factors
which depend on the pion mass.
\end{itemize}

With these two additional constraints we
can improve the form factors in the low $Q^2$ region
consistently with the chiral behavior,
in particular for the neutron electric form factor.
We also provide a direct connection
of the model with the chiral limit.

Next, we describe how the valence
quark and the pion cloud contributions
are integrated in the model.
The explicit parametrization for the pion cloud
and the connection with chiral perturbation theory
are presented later.

\subsection{Bare form factors}
\label{secBare}

In the covariant spectator quark model
a baryon $B$ is described as a system
with an off-mass-shell quark, free to
interact with photons, and two on-mass-shell quarks.
Integrating over the two on-mass-shell
quark momenta, we represent the quark pair as
an on-mass-shell diquark with an effective mass $m_D$,
and the baryon as a quark-diquark system~\cite{Nucleon,Nucleon2}.
This quark-diquark system is then described
by a transition vertex between the three-quark bound state
and the quark-diquark state,
that simulates  the
effect of confinement~\cite{Nucleon,Omega}.

The simplest representation for a
quark-diquark system with spin 1/2 and positive parity
is the S-wave configuration.
As in  \cite{Nucleon,OctetMag,OctetEMFF},
we represent the
wavefunction $\Psi_B$ for an octet baryon $B$
with momentum $P$ and the internal diquark momentum $k$.
In the S-wave approximation the wavefunction
is a combination of symmetric ($\ket{M_S}$) and anti-symmetric
($\ket{M_A}$) states in the quark pair (12),
and an S-wave radial (scalar) wavefunction  $\psi_B(P,k)$.
The explicit expressions can be found in  \cite{OctetEMFF}.

\subsubsection{Electromagnetic current.}

Taking into account that the wavefunction $\Psi_B$ is written
in terms of the wavefunctions
of a quark pair (12) and a single quark (3),
one can write the  electromagnetic
current associated with the baryon $B$
in impulse approximation \cite{Nucleon,Omega} as:
\ba
J_{0B}^\mu&=& 3 \sum_{\Gamma}
\int_k \overline \Psi_B (P_+,k)j_q^\mu \Psi_B (P_-,k),
\label{eqJB0}
\ea
where $j_q^\mu$ is the quark current operator,
$P_+$ ($P_-$) is the final (initial)
baryon momentum and $k$ the
momentum of the on-shell diquark.
$\Gamma= \left\{s, \lambda \right\}$ labels the
scalar diquark and the vectorial diquark
polarization $\lambda=0,\pm$.
The factor 3 in equation (\ref{eqJB0}) takes into account
the contributions for the current from the pairs $(13)$ and $(23)$,
where each pair has an identical contribution with that of the pair $(12)$.
The polarization indices are suppressed for simplicity.
The integral symbol stands for
\be
\int_k \equiv \int \frac{d^3 {\bf k}}{2 E_D (2\pi)^3},
\ee
where $E_D=\sqrt{m_D^2+ {\bf k}^2}$.

Generally, the baryon electromagnetic current~(\ref{eqJB0})
can be expressed as
\be
J_{0B}^\mu=
F_{10B}(Q^2) \, \gamma^\mu + F_{20B}(Q^2) \,
\frac{i \sigma^{\mu \nu}q_\nu}{2M_B},
\label{eqJB1}
\ee
where $F_{10B}$ and $F_{20B}$ are   respectively
the valence quark contributions for the
$F_{1B}(Q^2)$ and $F_{2B}(Q^2)$ form factors,
defined by equation (\ref{eqJ0}).
To simplify our notations we introduce
\ba
\tilde e_{0B} \equiv F_{10B}(Q^2), \hspace{.5cm}
\tilde \kappa_{0B} \equiv F_{20B}(Q^2),
\ea
where the tilde is included to remember
that these are functions of $Q^2$.
To represent these quantities for $Q^2=0$,
respectively the charge ($\tilde e_{0B}$)
and the anomalous magnetic moment ($\tilde \kappa_{0B}$)
we suppress the tildes as $e_{0B}$ and $\kappa_{0B}$.

\subsubsection{Quark electromagnetic current.}

The quark current operator $j_q^\mu$ has a
generic structure,
\be
j_q^\mu = j_1
\left(
\gamma^\mu - \frac{\not \! q q^\mu}{q^2}\right)
+ j_2 \frac{i \sigma^{\mu \nu} q_\nu}{2 M_N},
\label{eqJi}
\ee
where $M_N$ is the nucleon mass in vacuum
and $j_i$ ($i=1,2$) are $SU(3)$ flavor operators acting on the
third quark of the $\ket{M_A}$ or $\ket{M_S}$ state.
In the first term ${\not \! q} q^\mu / q^2$ is
included for completeness, but does not
contribute for elastic reactions.

The quark current $j_i$ ($i=1,2$) in equation (\ref{eqJi}),
can be decomposed as the sum of operators
acting on the quark 3 in $SU(3)$ flavor space~\cite{Omega},
\be
j_i=
\sfrac{1}{6} f_{i+} \lambda_0
+  \sfrac{1}{2}f_{i-} \lambda_3
+ \sfrac{1}{6} f_{i0} \lambda_s,
\ee
where
$\lambda_0=\mbox{diag}(1,1,0)$,
$\lambda_3=\mbox{diag}(1,-1,0)$
and  $\lambda_s=\mbox{diag}(0,0,-2)$
are the flavor space operators.
These operators act on
the quark wavefunction in flavor space,
$q=  (\begin{array}{c c c} \! u \, d \, s \!\cr
\end{array} )^T$.

The functions $f_{i\pm}(Q^2)$  ($i=1,2$) are normalized
by $f_{1\, n}(0)=1$ ($n=0,\pm$),
$f_{2\pm}(0)=\kappa_\pm$,
and $f_{20}(0)=\kappa_s$.
The isoscalar ($\kappa_+$) and isovector ($\kappa_-$)
anomalous magnetic moments
are defined in terms of the $u$ and $d$ quark
anomalous magnetic moments, $\kappa_+= 2 \kappa_u -\kappa_d$
and $\kappa_-= \sfrac{2}{3} \kappa_u+ \sfrac{1}{3} \kappa_d$.

\subsubsection{Dirac and Pauli bare form factors.}
\label{secFFbare}

To take into account the effect
of the states with mixed symmetry
in the baryon $B$ wavefunction ($\ket{M_A}$ and $\ket{M_S}$)
we sum over the quark flavors,
using  the coefficients~\cite{OctetMag,OctetEMFF,Omega},
\ba
j_i^A&=&
\bra{M_A} j_i \ket{M_A},
\label{eqjiA}\\
j_i^S&=&
\bra{M_S} j_i \ket{M_S},
\label{eqjiS}
\ea
for $i=1,2$.
The expressions for $j_i^A$ and $j_i^S$ ($i=1,2$)
are presented in
table \ref{tablePhiB} --see  \cite{OctetEMFF} for more details.

\begin{table}[t]
\caption{Mixed symmetric ($j_i^S$)
and antisymmetric ($j_i^A$)
coefficients for the octet baryons
appearing in equations (\ref{eqjiA}) and (\ref{eqjiS}).}
\label{tablePhiB}
\begin{center}
\begin{tabular}{l c c}
\hline
\hline
$B$   & $j_i^S$  &   $j_i^A$  \\
\hline
$p$     & $\sfrac{1}{6} (f_{i+}-f_{i-}) $ &
        $\sfrac{1}{6} (f_{i+}+3 f_{i-})  $ \\
$n$     & $\sfrac{1}{6} (f_{i+}+  f_{i-}) $ &
        $\sfrac{1}{6} (f_{i+}-  3 f_{i-})$ \\
\hline
$\Lambda$ & $\sfrac{1}{6}f_{i+}$ &
 $\sfrac{1}{18} (f_{i+}-  4 f_{i0})$ \\
\hline
$\Sigma^+$  & $\sfrac{1}{18}
(f_{i+} + 3 f_{i-} -4 f_{i0}) $ &
 $\sfrac{1}{6} (f_{i+}+3 f_{i-})  $ \\
$\Sigma^0$ &  $\sfrac{1}{36} (2 f_{i+}-  8 f_{i0})$
& $\sfrac{1}{6}f_{i+}$ \\
$\Sigma^-$ & $\sfrac{1}{18}
(f_{i+} - 3 f_{i-} -4 f_{i0}) $ &
        $\sfrac{1}{6} (f_{i+}-  3 f_{i-})           $ \\
\hline
$\Xi^0$ & $\sfrac{1}{18} (2 f_{i+} + 6  f_{i-} -2 f_{i0}) $ &
$-\sfrac{1}{3}f_{i0}$ \\
$\Xi^-$ & $\sfrac{1}{18} (2 f_{i+} -6  f_{i-} -2 f_{i0}) $ &
$-\sfrac{1}{3}f_{i0}$ \\
\hline
\hline
\end{tabular}
\end{center}
\end{table}

Using the
coefficients defined by equations (\ref{eqjiA}) and~(\ref{eqjiS})
we can write the spectator model
form factors for the octet baryons
characterized by a nucleon with mass $M_N$
and the baryon $B$ with mass $M_B$ as~\cite{OctetEMFF}
\ba
\tilde e_{0B} &=&
B(Q^2) 
\left(\frac{3}{2} j_1^A +
\frac{1}{2}
\frac{3 -\tau}{1+ \tau}
j_1^S - 2 \frac{\tau}{1+\tau} \frac{M_B}{M_N}
j_2^S \right),  \label{eqF1} \\
\tilde \kappa_{0B} &=&
B(Q^2) 
\left[
\left(
\frac{3}{2}
j_2^A
-\frac{1}{2} \frac{1-3\tau}{1+\tau} j_2^S \right) \frac{M_B}{M_N}
-2 \frac{1}{1+\tau} j_1^S
\right] , \nonumber \\
& &
\label{eqF2}
\ea
with $\tau=\sfrac{Q^2}{4M_B^2}$, and
\be
B(Q^2)= \int_k \psi_B(P_+,k) \psi_B(P_-,k),
\ee
the overlap integral between
the initial and final scalar wavefunctions.

The normalization of the wavefunction~\cite{Nucleon,OctetEMFF}
leads to $B(0)=1$.
Note that by construction the bare charge $e_{0B}$
and the dressed charge $e_B$ are the same, $e_{0B} \equiv e_B$.

We conclude then that the bare form factors
$\tilde e_{0B}$ and $\tilde \kappa_{0B}$
are determined by the $f_{i\,n}(Q^2)$ ($i=1,2$, $n=0,\pm$)
and $\psi_B(P,k)$.
The details of those parametrization are shown next.

We can also calculate the
quark core contributions for the
electric and magnetic form factors
using the following expressions:
\ba
& & G_{E0B}(Q^2)= \tilde e_B - \tau \, \tilde \kappa_B,
\label{eqGE0B}\\
& & G_{M0B}(Q^2)= \left[
\tilde e_B + \tilde \kappa_B \right] \frac{M_N}{M_B}.
\label{eqGM0B}
\ea
In equation (\ref{eqGM0B}) the factor
$\sfrac{M_N}{M_B}$ is included to be consistent
with the definition of $G_{MB}$.

Equations (\ref{eqF1}) and~(\ref{eqF2})
can be used either
in vacuum or in medium.
In medium we may just replace
the vacuum masses $M_N$ and $M_B$ respectively by
the effective masses $M_N^\ast$
and $M_B^\ast$.
The same procedure should also be carried out
in the coefficients
$j_i^A$ and $j_i^S$ ($i=1,2$),
namely, the vector meson masses in vacuum
should be replaced by those in medium.
As for equations (\ref{eqGE0B}) and~(\ref{eqGM0B})
the same rules are applied except that
the factor $\sfrac{M_N}{M_B}$ in vacuum
should be replaced by $\sfrac{M_N}{M_B^\ast}$
in medium (with the nucleon vacuum mass $M_N$)
according to our convention for $G_{MB}^\ast$.

\subsubsection{Parametrization of the quark current.}

To parameterize the quark current given by equation (\ref{eqJi}),
we adopt the structure inspired by the vector meson dominance (VMD)
mechanism as in  \cite{Nucleon,Omega},
\ba
& &
\hspace{-1cm}
f_{1 \pm} = \lambda_q
+ (1-\lambda_q)
\frac{m_\rho^2}{m_\rho^2+Q^2} + c_\pm \frac{M_h^2 Q^2}{(M_h^2+Q^2)^2},
\nonumber \\
& &
\hspace{-1cm}
f_{1 0} = \lambda_q
+ (1-\lambda_q)
\frac{m_\phi^2}{m_\phi^2+Q^2} + c_0 \frac{M_h^2 Q^2}{(M_h^2+Q^2)^2},
\nonumber \\
& &
\hspace{-1cm}
f_{2 \pm} = \kappa_\pm
\left\{
d_\pm  \frac{m_\rho^2}{m_\rho^2+Q^2} + (1-d_\pm)
\frac{M_h^2 }{M_h^2+Q^2} \right\}, \nonumber \\
& &
\hspace{-1cm}
f_{2 0} = \kappa_s
\left\{
d_0  \frac{m_\phi^2}{m_\phi^2+Q^2} + (1-d_0)
\frac{M_h^2}{M_h^2+Q^2}  \right\},
\label{eqQff}
\ea
where $m_\rho,m_\phi$ and $M_h$ are the masses respectively
corresponding to the light vector meson ($\rho$ meson),
the $\phi$ meson (associated with an $s \bar s$ state),
and an effective heavy meson
with mass $M_h= 2 M_N$ to represent
the short-range phenomenology.
We use the mass value $m_\rho$ for both isoscalar
(corresponding $\omega$ meson)
and isovector channels since
$m_\omega \simeq m_\rho$.
The coefficients $c_0,c_\pm$ and $d_0,d_\pm$ were
determined in the previous studies for nucleon
(model II)~\cite{Nucleon} and $\Omega^-$~\cite{Omega}.
The values are, respectively,
$c_+= 4.160$, $c_-= 1.160$, $d_+=d_-=-0.686$,
$c_0=4.427$ and $d_0=-1.860$~\cite{Omega}.
The constant $\lambda_q=1.21$ is obtained
so as to reproduce correctly the quark number
density in deep inelastic scattering~\cite{Nucleon}.

The quark form factors parameterized by
the VMD mechanism in equation (\ref{eqQff}) are particularly
convenient to extend the model to other regimes
besides the physical regime,
because the quark current is written in terms of the vector meson
and nucleon masses.
In the previous work the model was extended successfully
to the lattice QCD regime
replacing these masses by those of
the lattice regime~\cite{OctetEMFF}.
Furthermore, the model was also applied to
the lattice regime for the nucleon~\cite{Lattice},
$\gamma N \to \Delta$ reaction~\cite{LatticeD,Lattice},
and octet and decuplet baryons~\cite{OctetEMFF,Omega}.

\subsubsection{Scalar wavefunctions.}
\label{appSWF}

The scalar wavefunctions are given by~\cite{OctetEMFF},
\ba
\psi_N(P,k) &=&
\frac{N_N}{m_D(\beta_1 + \chi_{_N})(\beta_2+\chi_{_N})},
\label{eqPsiSN}\\
\psi_\Lambda(P,k) &=&
\frac{N_\Lambda}{m_D(\beta_1 + \chi_{_\Lambda})
(\beta_3+\chi_{_\Lambda})}, \\
\psi_\Sigma(P,k) &=&
\frac{N_\Sigma}{m_D(\beta_1 + \chi_{_\Sigma})
(\beta_3+\chi_{_\Sigma})}, \\
\psi_\Xi(P,k) &=&
\frac{N_\Xi}{m_D(\beta_1 + \chi_{_\Xi})
(\beta_4+\chi_{_\Xi})},
\label{eqPsiSXi}
\ea
where $N_B$
($B=N,\Lambda,\Sigma,\Xi$) are the normalization constants, and
\be
\chi_{_B} = \frac{(M_B-m_D)^2-(P-k)^2}{M_B m_D}.
\ee
Note that, except for the masses, the
$\Lambda$ and $\Sigma$ scalar wavefunctions are the same.
The normalization constants $N_B$
are determined by
\be
\int_k | \psi_B(\bar P,k)|^2=1,
\ee
where $\bar P=(M_B,0,0,0)$ is the baryon four-momentum at
its rest frame.

In equations (\ref{eqPsiSN})-(\ref{eqPsiSXi}) the parameters
$\beta_i$ ($i=1,..,4$) define the momentum range in units of $m_D$.
The parameter $\beta_1$
is associated with the long-range scale
(low-momentum range) that is common to all the octet baryon members.
As for the remaining
parameters $\beta_2,\beta_3,\beta_4 > \beta_1$,
they are associated
with the shorter range scale (larger momentum range).
Namely, $\beta_2$ defines the short-range scale
for the systems with only light quarks $u$ and $d$,
$\beta_3$ defines the short-range scale for
the systems with one strange quark, and
$\beta_4$ defines the scale for the systems
with two strange quarks.
As the strange quarks are heavier
than the $u$ and $d$ quarks, and
therefore more confined in the space,
we expect that $\beta_2 > \beta_3 > \beta_4$.
The parameters $\beta_i$ ($i=1,..,4$)
as well as
$\kappa_u$, $\kappa_d$ will be fixed later.

\subsection{Pion cloud dressing}
\label{secPionCloud}

We discuss here the pion cloud contributions
for the electromagnetic current and form factors
represented by the diagrams
in figure \ref{figPionCloud}.
Following  \cite{OctetMag}, we assume
the pion as the dominant meson excitation
to be included in the octet baryon form factors.
Then, the meson cloud contributions for
the octet baryon electromagnetic form factors can be described
in terms of 6 independent functions of $Q^2$,
related to the pion-baryon Feynman integral
as will be described next.

\subsubsection{Pion cloud electromagnetic currents.}

The pion cloud corrections, namely
the coupling of the photon
to the pion $J_\pi^\mu$, and
the coupling to the intermediate baryons
$J_{\gamma B}^\mu$,
defined by equation (\ref{eqJdecomp}),
can be written~\cite{OctetMag,OctetEMFF}
\ba
J_{\pi}^\mu
&=&
\left(
\tilde B_1 \gamma^\mu + \tilde B_2
\frac{i \sigma^{\mu \nu}q_\nu}{2M_{B}}
\right) G_{\pi B}, \label{eqJpi}\\
J_{\gamma B}^\mu &=&
\left(
\tilde C_1 \gamma^\mu + \tilde C_2
\frac{i \sigma^{\mu \nu}q_\nu}{2M_{B}} \right) G_{e B}+ \nonumber \\
& &
\left(
\tilde D_1 \gamma^\mu + \tilde D_2
\frac{i \sigma^{\mu \nu}q_\nu}{2M_{B}}\right)  G_{\kappa B}.
\label{eqJgB}
\ea
In the above, $\tilde B_i$, $\tilde C_i$ and $\tilde D_i$
($i=1,2$) are functions of $Q^2$ and
$G_{\pi B}$,
$G_{e B}$, $G_{\kappa B}$ are coefficients
that depend on the baryon species ($B=N,\Sigma,\Lambda,\Xi$).
We assume that the functions $\tilde B_i$, $\tilde C_i$
and $\tilde D_i$ are only weakly dependent on the baryon masses,
and the same for all the octet baryons as in  \cite{OctetMag}.
That allows a description of the pion cloud dressing
with a reduced number of coefficients.
We write $B_i$, $C_i$, and $D_i$ to represent respectively
the functions $\tilde B_i$, $\tilde C_i$ and $\tilde D_i$
at $Q^2=0$.

The coefficients $G_{\pi B}$,
$G_{e B}$, $G_{\kappa B}$ include the dependence
on the pion-baryon coupling constants.
According to $SU(3)$ symmetry~\cite{GellMann62,Carruthers}
the coupling constant of the pion ($\pi$) and baryons ($B$ and $B^\prime$),
$g_{\pi BB^\prime}$, can
be represented in terms of the ratio,
$\alpha=\frac{D}{F+D}$ and a global coupling constant
$g = g_{\pi NN}$, the $\pi NN$ coupling constant.
We can then express
$G_{\pi B}$, $G_{e B}$, $G_{\kappa B}$
in terms of the parameter $\alpha$
and a global factor $g^2$.
For convenience we absorb the factor $g^2$
in the functions $\tilde B_i,\tilde C_i$ and $\tilde D_i$ ($i=1,2$)
and represent the effect of the coupling
in terms of 4 independent constants~\cite{OctetMag,OctetEMFF}
associated with the octet baryon
species\footnote{The coefficient $\beta_N$
was not considered explicitly in the previous
works~\cite{OctetMag,OctetEMFF}, where $\beta_N \equiv 1$,
but it is included here for completeness and to clarify
the extension of the model to the in-medium case.},
\ba
& &
\beta_N= 1, \label{EqbetaN} \\
& &
\beta_\Lambda = \frac{4}{3} \alpha^2,
\label{EqbetaL} \\
& &
\beta_\Sigma= 4(1-\alpha)^2,
\label{EqbetaS} \\
& &
\beta_\Xi = (1-2 \alpha)^2.
\label{EqbetaX}
\ea
These constants encapsulate the effect of the coupling constants.
The explicit dependence of $G_{\pi B}$, $G_{e B}$ and  $G_{\kappa B}$
on the constants given by equations (\ref{EqbetaN})-(\ref{EqbetaX})
and on the bare form factors $\tilde e_{0B}$ and $\tilde \kappa_{0B}$
was derived in  \cite{OctetMag,OctetEMFF}.
In the following we use the
results of  \cite{OctetEMFF}.

From the equations above, we get
$\beta_N=1$, $\beta_\Lambda= 0.48$, $\beta_\Sigma=0.64$
and $\beta_\Xi=0.04$ with $\alpha=0.6$,
determined in combination with an $SU(6)$ quark model.
It is therefore expected that the pion cloud contributions
are small for the $\Xi$ system.
In this case the kaon cloud contribution may be more 
significant.

\subsubsection{Dressed form factors.}
\label{appGdressed}

The octet baryon dressed form factors
associated with the current~(\ref{eqJdecomp}),
are obtained by
the contributions from the quark core
given by equations (\ref{eqF1})-(\ref{eqF2}), and
the pion cloud dressing via equations (\ref{eqJpi})-(\ref{eqJgB}):
\ba
J_B^\mu  &=& 
Z_B
\left[
\tilde e_{0B} + G_{\pi B} \tilde B_1 +
G_{e B} \tilde C_1 + G_{\kappa B} \tilde D_1 \right] \gamma^\mu +
\nonumber \\
& &
Z_B
\left[
\tilde \kappa_{0B} + G_{\pi B} \tilde B_2 +
G_{e B} \tilde C_2 + G_{\kappa B} \tilde D_2 \right]
\frac{i \sigma^{\mu \nu} q_\nu}{2 M_B}. \nonumber \\
\label{eqJtot}
\ea
Using the above expressions
and the definition of the form factors~(\ref{eqJ0}),
we can write down the final results for the
form factors $F_{1B}$ and $F_{2B}$:
\ba
F_{1p} &=& Z_N
\left\{
\tilde e_{0p} + 2 \beta_N \tilde B_1 +
\beta_N (\tilde e_{0p} + 2 \tilde e_{0n}) \tilde C_1
+ \beta_N (\tilde \kappa_{0p} + 2 \tilde \kappa_{0n})\tilde D_1 \sfrac{}{} \!
\right\},  \label{eqF1p} \\
F_{1n} &=&
Z_N
\left\{ \tilde e_{0n} - 2 \beta_N \tilde B_1 +
\beta_N (2 \tilde e_{0p} + \tilde e_{0n}) \tilde C_1
+ \beta_N (2\tilde \kappa_{0p} + \tilde \kappa_{0n})\tilde D_1 \sfrac{}{} \!
\right\}, \label{eqF1n} \\
& & \nonumber \\
F_{1 \Lambda} &=&
Z_\Lambda
\left\{ \tilde e_{0 \Lambda}  +
\beta_\Lambda(\tilde e_{0 \Sigma^+} + \tilde e_{0\Sigma^0}
+ \tilde e_{0 \Sigma^-}) \tilde C_1
+ \beta_\Lambda (\tilde \kappa_{0 \Sigma^+} + \tilde \kappa_{0 \Sigma^0}
+ \tilde \kappa_{0 \Sigma^-})\tilde D_1 \sfrac{}{} \!
\right\},  \nonumber \\
& &\label{eqF1Lambda}\\
& & \nonumber \\
F_{1 \Sigma^+} &=&
Z_\Sigma
\left\{
\tilde e_{0 \Sigma^+}  + (\beta_\Sigma + \beta_\Lambda) \tilde B_1
+ \left[
\beta_\Sigma(\tilde e_{0 \Sigma^+} + \tilde e_{0\Sigma^0})
+ \beta_\Lambda \tilde e_{0 \Lambda} \right] \tilde C_1  \frac{}{} \!
\right.
\nonumber \\
& & \left. +
\left[
\beta_\Sigma(\tilde \kappa_{0 \Sigma^+} + \tilde \kappa_{0\Sigma^0})
+ \beta_\Lambda \tilde \kappa_{0 \Lambda} \right] \tilde D_1
 \frac{}{} \!
\right\}, \\
F_{1 \Sigma^0} &=&
Z_\Sigma
\left\{
\tilde e_{0 \Sigma^0}
+
\left[
\beta_\Sigma(\tilde e_{0 \Sigma^+} + \tilde e_{0\Sigma^-})
+ \beta_\Lambda \tilde e_{0 \Lambda} \right] \tilde C_1  \frac{}{} \!
\right.
\nonumber \\
& & \left. +
\left[
\beta_\Sigma(\tilde \kappa_{0 \Sigma^+} + \tilde \kappa_{0\Sigma^-})
+ \beta_\Lambda \tilde \kappa_{0 \Lambda} \right] \tilde D_1
 \frac{}{} \!
\right\}, \\
F_{1 \Sigma^-} &=&
Z_\Sigma
\left\{
\tilde e_{0 \Sigma^-} - (\beta_\Sigma+ \beta_\Lambda) \tilde B_1
+
\left[
\beta_\Sigma(\tilde e_{0 \Sigma^0} + \tilde e_{0\Sigma^-})
+ \beta_\Lambda \tilde e_{0 \Lambda} \right] \tilde C_1  \frac{}{} \!
\right.
\nonumber \\
& & \left. +
\left[
\beta_\Sigma(\tilde \kappa_{0 \Sigma^0} + \tilde \kappa_{0\Sigma^-})
+ \beta_\Lambda \tilde \kappa_{0 \Lambda} \right] \tilde D_1
 \frac{}{} \!
\right\}, \\
& & \nonumber \\
F_{1 \Xi^0} &=&
Z_\Sigma
\left\{
\tilde e_{0 \Xi^0} + 2 \beta_\Xi \tilde B_1
+
\beta_\Xi(\tilde e_{0 \Xi^0} + 2 \tilde e_{0\Xi^-}) \tilde C_1 \frac{}{} \!
\right.
\nonumber \\
& & \left. +
\beta_\Xi(\tilde \kappa_{0 \Xi^0} + 2 \tilde \kappa_{0\Xi^-}) \tilde D_1
 \frac{}{} \!
\right\}, \\
F_{1 \Xi^-} &=&
Z_\Sigma
\left\{
\tilde e_{0 \Xi^-} - 2 \beta_\Xi \tilde B_1
+
\beta_\Xi(2 \tilde e_{0 \Xi^0} +  \tilde e_{0\Xi^-}) \tilde C_1
\frac{}{} \! \right. \nonumber \\ & & \left.
+
\beta_\Xi(2 \tilde \kappa_{0 \Xi^0} + \tilde \kappa_{0\Xi^-}) \tilde D_1
 \frac{}{} \!
\right\},
\label{eqF1XiM}
\ea

\ba
F_{2p}&=&
Z_N
\left\{ \tilde \kappa_{0p} + 2 \beta_N \tilde B_2
 +
\beta_N (\tilde e_{0p} + 2\tilde e_{0n}) \tilde C_2
+ \beta_N (\tilde \kappa_{0p} + 2 \tilde \kappa_{0n})\tilde D_2 \sfrac{}{} \!
\right\},
\label{eqF2p} \\
F_{2n}&=&
Z_N
\left\{ \tilde \kappa_{0n} - 2 \beta_N \tilde B_2
   + \beta_N
(2 \tilde e_{0p} +  \tilde e_{0n}) \tilde C_2
+ \beta_N (2 \tilde \kappa_{0p} +  \tilde \kappa_{0n})\tilde D_2 \sfrac{}{} \!
\right\},  \label{eqF2n} \\
&& \nonumber \\
F_{2 \Lambda} &=&
Z_\Lambda
\left\{ \tilde \kappa_{0 \Lambda}  +
\beta_\Lambda(\tilde e_{0 \Sigma^+} + \tilde e_{0\Sigma^0}
+ \tilde e_{0 \Sigma^-}) \tilde C_2
+ \beta_\Lambda (\tilde \kappa_{0 \Sigma^+} + \tilde \kappa_{0 \Sigma^0}
+ \tilde \kappa_{0 \Sigma^-})\tilde D_2  \sfrac{}{} \!
\right\}, \nonumber \\
& & \label{eqF2Lambda}
\ea
\ba
F_{2 \Sigma^+} &=&
Z_\Sigma
\left\{ \frac{}{} \!
\tilde \kappa_{0 \Sigma^+}  + (\beta_\Sigma + \beta_\Lambda) \tilde B_2
+
\left[
\beta_\Sigma(\tilde e_{0 \Sigma^+} + \tilde e_{0\Sigma^0})
+ \beta_\Lambda \tilde e_{0 \Lambda} \right] \tilde C_2  \frac{}{} \!
\right.
\nonumber \\
& & \left. +
\left[
\beta_\Sigma(\tilde \kappa_{0 \Sigma^+} + \tilde \kappa_{0\Sigma^0})
+ \beta_\Lambda \tilde \kappa_{0 \Lambda} \right] \tilde D_2
 \frac{}{} \!
\right\}, \\
F_{2 \Sigma^0} &=&
Z_\Sigma
\left\{ \frac{}{} \!
\tilde \kappa_{0 \Sigma^0}
+
\left[
\beta_\Sigma(\tilde e_{0 \Sigma^+} + \tilde e_{0\Sigma^-})
+ \beta_\Lambda \tilde e_{0 \Lambda} \right] \tilde C_2  \frac{}{} \!
\right.
\nonumber \\
& & \left. +
\left[
\beta_\Sigma(\tilde \kappa_{0 \Sigma^+} + \tilde \kappa_{0\Sigma^-})
+ \beta_\Lambda \tilde \kappa_{0 \Lambda} \right] \tilde D_2
 \frac{}{} \!
\right\}, \\
F_{2 \Sigma^-} &=&
Z_\Sigma
\left\{
\tilde \kappa_{0 \Sigma^-}   - (\beta_\Sigma+ \beta_\Lambda) \tilde B_2
+
\left[
\beta_\Sigma(\tilde e_{0 \Sigma^0} + \tilde e_{0\Sigma^-})
+ \beta_\Lambda \tilde e_{0 \Lambda} \right] \tilde C_2  \frac{}{} \!
\right.
\nonumber \\
& & \left. +
\left[
\beta_\Sigma(\tilde \kappa_{0 \Sigma^0} + \tilde \kappa_{0\Sigma^-})
+ \beta_\Lambda \tilde \kappa_{0 \Lambda} \right] \tilde D_2
 \frac{}{} \!
\right\}, \\
&& \nonumber \\
F_{2 \Xi^0} &=&
Z_\Sigma
\left\{
\tilde \kappa_{0 \Xi^0} + 2 \beta_\Xi \tilde B_2
+
\beta_\Xi(\tilde e_{0 \Xi^0} + 2 \tilde e_{0\Xi^-}) \tilde C_2
\frac{}{} \! \right. \nonumber \\ & & \left.
+
\beta_\Xi(\tilde \kappa_{0 \Xi^0} + 2 \tilde \kappa_{0\Xi^-}) \tilde D_2
 \frac{}{} \!
\right\}, \\
F_{2 \Xi^-} &=&
Z_\Sigma
\left\{
\tilde \kappa_{0 \Xi^-} - 2 \beta_\Xi \tilde B_2
+
\beta_\Xi(2 \tilde e_{0 \Xi^0} +  \tilde e_{0\Xi^-}) \tilde C_2 \frac{}{} \!
\right.
\nonumber \\
& & \left. +
\beta_\Xi(2 \tilde \kappa_{0 \Xi^0} + \tilde \kappa_{0\Xi^-}) \tilde D_2
 \frac{}{} \!
\right\}.
\label{eqF2XiM}
\ea

The normalization constants are
\ba
& & Z_N= \Big[1+3\beta_N B_1\Big]^{-1} \nonumber \\
& &Z_\Lambda=\Big[1+3\beta_\Lambda B_1\Big]^{-1}, \nonumber \\
&&Z_\Sigma=\bigg[1+\Big(2\beta_\Sigma + \beta_\Lambda \Big)B_1\bigg]^{-1},
\nonumber \\
&&Z_\Xi=\Big[1+3\beta_\Xi B_1\Big]^{-1},
\label{eqZB}
\ea
and we refer to  \cite{OctetMag,OctetEMFF} for more details.

Using the above expressions,
one can calculate
the electric and magnetic form factors
both in vacuum, based on equations (\ref{eqGEB0}) and~(\ref{eqGMB0}),
and in medium based on equations (\ref{eqGEm})-(\ref{eqGMm}).

From the discussions in the previous sections,
the baryon form factors
$F_{iB}$ ($i=1,2$) may be decomposed into
\be
F_{iB}(Q^2)= Z_B \left[  F_{i0B}(Q^2) + \delta F_{iB} (Q^2)\right],
\ee
where $F_{10B}$ and $F_{20B}$
are defined by equation (\ref{eqJB1}),
and the corresponding pion cloud contributions
$\delta F_{1 B}$ and $\delta F_{2 B}$
are given in equations (\ref{eqF1p})-(\ref{eqF2XiM}).
It is natural to regard $Z_{B} F_{i0B}$ as representing
the valence quark effects and $Z_{B} \delta F_{iB}$
those of the pion cloud.

The same decomposition can be applied for
the electric and magnetic form factors:
\ba
\hspace{-1.cm}
& &
G_{E B}(Q^2)= Z_B \left[ G_{E0B}(Q^2) + \delta G_{E B}(Q^2)\right], \nonumber \\
\hspace{-1.cm}
& &
G_{M B}(Q^2)= Z_B \left[ G_{M0B} (Q^2)+ \delta G_{M B}(Q^2)\right],
\label{eqGX1}
\ea
where $G_{E0B}, G_{M0B}$ are defined
by equations (\ref{eqGE0B}) and~(\ref{eqGM0B}), and
$\delta G_{E B}= \delta F_{1B} -\tau \delta F_{2B}$ and
$\delta G_{M B}= \left[\delta F_{1B} + \delta F_{2B} \right]\sfrac{M_N}{M_B}$.
In this case $Z_B \delta G_{EB}$, and $Z_B \delta G_{MB}$ reflect
the dressing of the pion cloud.
To estimate the pion cloud contributions,
we compare the full result,
$G_{EB}$ or $G_{MB}$, with the total contributions
of the valence quark core,
$Z_B G_{E0B}$ or $Z_B G_{M0B}$.
The difference is
the pion cloud contributions, $Z_B \delta G_{EB}$ or $Z_B \delta G_{MB}$.

\section{Including chiral symmetry in the pion cloud parametrization}

The effect of chiral symmetry in the
electromagnetic structure of the octet baryons
can be analyzed by studying the dependence
of the form factors on $m_\pi$ and
the corresponding behavior in
the chiral limit ($m_\pi \to 0$).
In the chiral limit the pion cloud extends
to infinity, leading to the divergence
of the nucleon radius.

In this section we start reviewing
the main features of the chiral behavior
for the nucleon radii.
At this stage we do not attempt to describe
the nucleon magnetic moments
including the pion mass dependence,
since the dependence is milder than
that for the nucleon radii~\cite{Leinweber99,Young04,Hall12}.
Next, we derive the general expressions
for the nucleon radii in the present model.
Finally, we present the newly updated
parametrization for the pion cloud contributions,
and describe how the chiral behavior
is implemented in the model.

\subsection{Nucleon radii in the chiral limit}

The leading order effects of chiral symmetry
can be better observed in the
nucleon isovector form factor~\cite{Hammer04}
defined by,
\ba
&&
F_1^V(Q^2)= F_{1p}(Q^2)-F_{1n}(Q^2),
\label{eqF1v}\\
&&
F_2^V(Q^2)= F_{2p}(Q^2)-F_{2n}(Q^2).
\label{eqF2v}
\ea
Using these decompositions
we can define the isovector
Dirac and Pauli square radii:
\ba
& &
(r_1^V)^2= - 6 \left.
\frac{d F_1^V}{d Q^2} \right|_{Q^2=0} \; , \\
& &
(r_2^V)^2= - 6 \left.
\frac{d F_2^V}{d Q^2} \right|_{Q^2=0} \; .
\ea
Note that in the definitions we
do not normalize the respective radius
at $Q^2=0$ as usually done.
To compare $(r_2^V)^2$ with the experimental value
the result must be divided by $\kappa_V = \kappa_p -\kappa_n \simeq 3.7$
(isovector anomalous magnetic moment).

We now discuss the expected result
in the small pion mass limit.
According to $\chi$PT~\cite{Bernard95,Perdrisat07}
the isovector square radii $(r_1^V)^2$ and $(r_2^V)^2$
can be expressed as
\ba
& &
(r_1^V)^2=
- \frac{\alpha_1}{\alpha_0} \log m_\pi + A_1,
\label{eqR1b}\\
& &
(r_2^V)^2=
+ \frac{\alpha_2}{\alpha_0} \frac{M}{m_\pi} + A_2,
\label{eqR2b}
\ea
where $m_\pi$ and $M$ are the pion and nucleon masses,
respectively, and
\ba
& &
\alpha_0= 8 \pi^2 F_\pi^2, \\
& &
\alpha_1= 5 g_A^2 + 1, \\
& &
\alpha_2= \pi g_A^2,
\ea
with $F_\pi= 93$ MeV and $g_A=1.27$.
In the above, $A_1$ and $A_2$
represent constant terms and higher powers of $m_\pi^2$,
that can be expressed by constants
at the physical point.

We take a pragmatic approach and fix these constant
values using the experimental values of
$(r_1^V)^2$ and $(r_2^V)^2$.
The results obtained from the average values\footnote{For
the neutron radii $r_{En}^2$ and $r_{Mn}^2$
we use the results suggested by PDG~\cite{PDG}.
For the proton radii there is still controversy.
For $r_{Mp}^2$ we take the arithmetic
average of the 3 results listed.
As for $r_{Ep}^2$
we simply take the arithmetic
average of the first 12 results from the list (since 2000),
irrespective of the different types of the determinations,
electronic, muonic, or others.}
of the PDG results~\cite{PDG} are presented in table \ref{tabRadiiN}.
From the table we find
$A_1\simeq 12.05$ fm$^2$ and
$A_2 \simeq -38.20$ fm$^2$.
These results are obtained assuming the physical
value of $m_\pi$ and the experimental results.

\begin{table}[t]
\caption{Nucleon radii.
We adopt the average values from the PDG~\cite{PDG}
more recent results.
All values in fm$^2$.
The Pauli isovector square radii normalized
is $\sfrac{(r_2^V)^2}{\kappa_V}=0.418\pm 0.012$ fm$^2$.}
{\footnotesize
\begin{center}
\begin{tabular}{c c c c c c}
\hline
\hline
$r_{Ep}^2$ & $r_{Mp}^2$ & $r_{En}^2$ & $r_{Mn}^2$  &
$(r_1^V)^2$ & $(r_2^V)^2$ \\
\hline
0.7634$\pm$0.0140   & 0.6983$\pm$0.0109   &
$-$0.1161$\pm$0.0022    &  0.7430$\pm$0.0134    &
0.0634$\pm$0.0142 & 1.5509$\pm$0.0430  \\
\hline
\hline
\end{tabular}
\end{center} }
\label{tabRadiiN}
\end{table}

\subsection{Isovector form factors and respective radii}
\label{secIsoRadius}

We discuss now the isovector form factors
given by the present model.

Using equations (\ref{eqF1p})-(\ref{eqF1n})
and (\ref{eqF2p})-(\ref{eqF2n}) we can write
\ba
& &
F_1^V(Q^2)=
Z_N \left\{ (\tilde e_{0p} -\tilde e_{0n})
+ 4 \tilde B_1
- (\tilde e_{0p} -\tilde e_{0n}) \tilde C_1
-(\tilde \kappa_{0p} -\tilde \kappa_{0n}) \tilde D_1
\right\}, \\
& &
F_2^V(Q^2) =
Z_N \left\{
(\tilde \kappa_{0p} -\tilde \kappa_{0n})  \frac{}{}
+ 4\tilde B_2 -  (\tilde e_{0p} -\tilde e_{0n}) \tilde C_2
- (\tilde \kappa_{0p} -\tilde \kappa_{0n}) \tilde D_2
\right\}.
\label{eqF2V}
\ea
From these equations we conclude
that the nucleon isovector form factors
are given by the difference
between the proton and neutron bare
form factors ($\tilde e_{0N}$ or $\tilde \kappa_{0N}$)
and the pion cloud contributions.

The results for the isovector square radii are given by
\ba
& &
(r_1^V)^2=
Z_N \left\{
- 24 \left. \frac{d \tilde B_1}{d Q^2} \right|_{0} +
R_1
\right\},
\label{eqR1a} \\
& &
(r_2^V)^2=
Z_N \left\{
- 24 \left. \frac{d \tilde B_2}{d Q^2} \right|_{0} +
R_2
\right\},
\label{eqR2a}
\ea
where
\ba
R_1&=&
6 \left. \frac{d \tilde C_1}{d Q^2} \right|_{0} +
(1-C_1) (r_{10p}^2- r_{10n}^2)
+ 6 (\kappa_{0p} -\kappa_{0n})
\left. \frac{d \tilde D_1}{d Q^2} \right|_{0},
\label{eqR1}\\
R_2&=&
 6 \left. \frac{d \tilde C_2}{d Q^2} \right|_{0}
 -  (r_{10p}^2- r_{10n}^2) C_2
+ (1-D_2) (r_{20p}^2- r_{20n}^2)  \nonumber \\
& &+ 6 (\kappa_{0p} -\kappa_{0n})
\left. \frac{d \tilde D_2}{d Q^2} \right|_{0}.
\label{eqR2}
\ea
In the above expressions
$\left. \right|_0$
stands for the derivative at $Q^2=0$, and
\ba
r_{i0B}^2= - 6 \left. \frac{d F_{i0B}}{d Q^2} \right|_{Q^2=0} (i=1,2),
\ea
represent the {\it bare} radii.

The expressions (\ref{eqR1a}) and~(\ref{eqR2a})
are still general.
We now discuss the constraints of
chiral perturbation theory for the model.

\subsection{Pion cloud parametrization}
\label{secPionCloud2}

We consider the following
parametrizations for the functions $\tilde B_1$
and $\tilde B_2$:
\ba
\tilde B_1 &=&
B_1 \left( \frac{\Lambda_1^2}{\Lambda_1^2 + Q^2}
\right)^5
\left[
1 + \frac{1}{Z_N B_1}
\left(
\frac{1}{24} \frac{\alpha_1}{\alpha_0} \log m_\pi
+ b_1^\prime
\right)
Q^2
\right],
\label{eqB1}
\\
\tilde B_2&=&
B_2 \left( \frac{\Lambda_2^2}{\Lambda_2^2 + Q^2}
\right)^6 
\left[
1 + \frac{1}{Z_N B_2}
\left(
-\frac{1}{24} \frac{\alpha_2}{\alpha_0} \frac{M}{m_\pi}
+ b_2^\prime
\right)
Q^2
\right].
\label{eqB2}
\ea
Here $B_1$ and $B_2$
are constants given respectively by $\tilde B_1(0)$ and $\tilde B_2(0)$,
and $\Lambda_1,\Lambda_2$ are two cutoffs
to be fixed by a fit to the data.

As for $b_1^\prime$ and $b_2^\prime$
they are two additional parameters that will
be fixed by the
experimental results for the nucleon isovector square radii,
equations (\ref{eqR1b}) and~(\ref{eqR2b}).

Inserting the above expressions into
equations (\ref{eqR1a}) and~(\ref{eqR2a}) we can write
\ba
& &
(r_1^V)^2=
 - \frac{\alpha_1}{\alpha_0} \log m_\pi
+ Z_N
\left\{
-24 \frac{b_1^\prime}{Z_N}
- 120 \frac{B_1}{\Lambda_1^2}
 + R_1
\right\}, \nonumber \\
& &
\label{eqR1f}
\\
& &
(r_2^V)^2=
 +  \frac{\alpha_2}{\alpha_0} \frac{M}{m_\pi}
+ Z_N
\left\{
- 24\frac{b_2^\prime}{Z_N}
- 144 \frac{B_1}{\Lambda_1^2}
 + R_2
\right\}. \nonumber \\
& &
\label{eqR2f}
\ea
Note that the second term in each equation
above should be identified with $A_1$ and $A_2$
in equations (\ref{eqR1b}) and~(\ref{eqR2b}), respectively.

As for the remaining functions
we use
 \ba
& & \tilde C_1= B_1\left(\frac{\Lambda_{1}^2}{\Lambda_{1}^2 + Q^2} \right)^2,
\label{eqC1}
\\
& & \tilde D_1= D_1^\prime \frac{Q^2\Lambda_{1}^4}{(\Lambda_{1}^2 + Q^2)^3},
\\
& & \tilde C_2= C_2\left(\frac{\Lambda_{2}^2}{\Lambda_{2}^2 + Q^2} \right)^3,
\\
& & \tilde D_2= D_2\left(\frac{\Lambda_{2}^2}{\Lambda_{2}^2 + Q^2} \right)^3.
\label{eqD2}
\ea
Again $C_2$ and $D_2$ are the constants
given by the value at $Q^2=0$ for the
respective functions.
$D_1^\prime$ is a new constant
defined by $D_1^\prime = \sfrac{1}{\Lambda_{1}^2}\sfrac{d D_1}{d Q^2}(0)$.
Note that $\tilde C_1(0)= B_1$, a constraint
required by the conditions $F_{1p}(0)=1$ and
$F_{1n}(0)=0$ (nucleon charges)~\cite{OctetMag,OctetEMFF}.
The definition of $\tilde D_1$ that vanishes
at $Q^2=0$ is also motivated by the
nucleon charge conditions.
The parametrizations used for $\tilde C_1, \tilde D_1, \tilde C_2$
and $\tilde D_2$ are the same as
those presented in  \cite{OctetEMFF}.
The leading order chiral effects in the
form factors in the present parametrization come
exclusively from
$\tilde B_1$ and $\tilde B_2$.
We choose to use the same cutoffs in $C_i,D_i$
as those used in $B_i$ ($\Lambda_1$ and $\Lambda_2$,
for the Dirac and Pauli form factors)
in order to reduce the number of parameters in the model~\cite{OctetEMFF}.

Finally, we can now write down,
\ba
R_1&=&
-12 \frac{B_1}{\Lambda_1^2}
+ (1-C_1) (r_{10p}^2- r_{10n}^2) 
+ 6 (\kappa_{0p}-\kappa_{0n}) \frac{D_1^\prime}{\Lambda_1^2}, \\
R_2&=&
-18 \frac{C_2}{\Lambda_2^2}
+ (1-D_2) (r_{20p}^2- r_{20n}^2)
\nonumber \\ & &
-(r_{10p}^2- r_{10n}^2) C_2
-18 (\kappa_{0p}-\kappa_{0n}) \frac{D_2}{\Lambda_2^2}.
\ea
The values of the
coefficients $b_1^\prime$ and $b_2^\prime$
can now be determined
comparing equations (\ref{eqR1f}) and~(\ref{eqR2f})
with equations (\ref{eqR1b}) and~(\ref{eqR2b}).

The choice of the powers included in the
pion cloud functions is phenomenological
and motivated by the expected falloff
of the quark-antiquark contributions
in the large $Q^2$ limit~\cite{Brodsky75}
as well as the magnitude of the pion cloud
contributions estimated for the
$\gamma N \to \Delta$ reaction~\cite{NDelta,NDeltaD,LatticeD}.
With the present parametrization the
pion cloud contributions fall off by a factor $1/Q^4$
faster than the falloff of the valence quark contributions.

\section{Extension of the spectator quark model to the in-medium regime}
\label{secInMediumRegime}

We now discuss the extension of the
model for the in-medium regime.
In general we consider the modifications
of the model due to the in-medium environment.
As for the valence quark core part,
the in-medium hadron masses appearing in the model,
$M_B$ ($B=N,\Lambda,\Sigma,\Xi$),
$m_\rho$, $m_\phi$ and $M_h$,
will be respectively denoted by
$M_B^*$ ($B=N,\Lambda,\Sigma,\Xi$), $m_\rho^\ast$, $m_\phi^\ast$  and
$M_h^\ast$ (given by $2M_N^*$).
On the other hand, for the pion cloud effects
we consider the modifications of the
pion-baryon couplings in the in-medium regime
as will be explained next.
In this study we do not include any final state interactions
nor meson exchange current, as mentioned in introduction.

\subsection{In-medium regime: quark core}
\label{InMediumCore}

Although lattice QCD simulation has been very rapidly developing recently,
it is still very difficult to study the properties of hadrons in
finite nucleon (baryon) densities near and higher than the normal
nuclear matter densities.
Thus, we need to resort to some phenomenological models which
have proven successful in studying nuclear phenomena and nuclear processes
based on the quark degrees of freedom.
For this purpose, we use the quark-meson coupling
(QMC) model~\cite{QMC,QMCReview}, which has been successfully
applied to study the properties of baryons and mesons in
nuclei and nuclear medium.

We note that, combined with the cloudy bag model
(CBM)~\cite{CBMTony},
QMC had indeed predicted~\cite{QMCEMFFMedium} an in-medium modification
of the bound proton
electromagnetic form factors which turned out to
be consistent with the experimentally
observed modification~\cite{Dieterich01,Strauch03,Paolone10,Malace11}.
Below, we use a different model,
the covariant spectator quark model~\cite{Nucleon,FixedAxis,NDelta},
which was successfully applied
to study the octet baryon electromagnetic
form factors~\cite{OctetEMFF} utilizing the lattice QCD simulation data.
By the use of different models, we hope to shed light on
the mechanism of the in-medium modification of the bound proton
electromagnetic form factors.
Furthermore, we predict in-medium electromagnetic form factors
of all members of the octet baryons.

The current $j_q^\mu$, given by equation (\ref{eqJi}), is
also characterized by the corresponding in-medium masses
in terms of the two components
$j_1$ and $j_2$, which are represented
based on the VMD parametrization in equations (\ref{eqQff}).

In the quark current~(\ref{eqJi}),
we replace the coefficient of the
Pauli form factor $1/(2 M_N)$
by $1/(2 M_N^*)$ in the in-medium regime.
As for the quark form factors,
we use equation (\ref{eqQff})
with the meson masses replaced by
the respective in-medium masses
similarly to the lattice regime studies~\cite{OctetEMFF,Omega,LatticeD,Lattice}.
That is, we replace $m_\rho$ and $m_\phi$
 by the in-medium $\rho$ mass $m_\rho^*$,
and $\phi$ mass $m_\phi^\ast$,
and the effective heavy meson mass of $M_h = 2 M_N$ by $2 M_N^*$.

As for the wavefunctions $\Psi_B$,
$M_B$ is replaced by $M_B^*$
in medium.
This applies for the radial (scalar) wavefunctions
(\ref{eqPsiSN})-(\ref{eqPsiSXi}).
As explained
the scalar wavefunctions for
the octet baryons are represented in terms of
four independent momentum range parameters $\beta_i$ ($i=1,..,4$).
We assume that these parameters are independent of
the baryon masses in vacuum, and therefore also
independent of the in-medium baryon masses.
There is no need to modify
the diquark mass in in medium,
since the electromagnetic form factors are
independent of it \cite{Nucleon,OctetEMFF}.
For the in-medium regime
the pion mass $m_\pi^*$, which is relevant for
the present study, it was estimated in
 \cite{fpiMedium} that $m_\pi^* \simeq m_\pi$ at normal
nuclear matter density (0.15 fm$^{-3}$).
Thus, we use the vacuum value, $m_\pi = 138.0$ MeV for
the densities considered in this study.
The in-medium hadron masses mostly calculated in QMC,
and the relevant values
for the calculation, are listed in table \ref{tableMass}.

Using the model extended to the in-medium regime, namely, using
the quark currents and baryon wavefunctions for the in-medium regime,
we can calculate the form factors
$G_{EB}^\ast$ and $G_{MB}^\ast$ in a nuclear medium
by the expressions given in section \ref{secBare},
for a model with no pion cloud dressing
with the in-medium masses calculated in QMC for a given
nuclear density.

\begin{table}[t]
\caption{Hadron masses (in MeV) necessary for the in-medium regime of the model
with $\rho_0=0.15$ fm$^{-3}$.
For the vacuum case ($\rho=0$)
we take $m_\rho$ as the average of $\rho$ and $\omega$ masses
as originally used in the model~\cite{Nucleon,OctetEMFF}.}
\begin{center}
\begin{tabular}{l c c c}
\hline
\hline
      &$\rho=0$ &$\rho=0.5 \,\rho_0$ &$\rho=\rho_0$ \\
\hline
$M_N$           &939.0  &831.3  &754.5  \\
$M_\Lambda$     &1116.0 &1043.9 &992.7  \\
$M_\Sigma$      &1192.0 &1121.4 &1070.4 \\
$M_\Xi$         &1318.0 &1282.2 &1256.7 \\
$m_\rho$        &779.0  & 706.1  &653.7  \\
$m_\phi$        &1019.5 &1019.1 &1018.9 \\
$m_\pi$         &138.0  &138.0  &138.0  \\
\hline
\hline
\end{tabular}
\end{center}
\label{tableMass}
\end{table}

\subsection{In-medium regime: pion cloud}
\label{InMediumCloud}

To extend the pion cloud effects of the model to the in-medium regime,
we need other ingredients.
Although we do not have a good control,
we estimate the effects of the pion cloud in the nuclear medium.
Thus, the estimates of the pion cloud effects in the nuclear medium
presented below, should be taken in caution.
Main task for this is to calculate the modifications of the
baryon-pion coupling constants in medium, $g_{\pi BB'}^*$
($B,B'=N,\Lambda,\Sigma,\Xi$).
[As before, we denote the properties in the nuclear medium
by asterisk $^\ast$.]
For the moment, we consider below the $\pi B B$ diagonal case, and omit the
$\pi\Lambda\Sigma$ coupling which appear in the pion cloud
effects in the $\Lambda$ and $\Sigma$ cases
containing the $g_{\pi \Lambda \Sigma}$ coupling
from the discussion,
but the procedure can be extended to also
for the $\pi \Lambda \Sigma$ case.
We assume that the pion cloud parametrization functions
given by equations (\ref{eqB1})-(\ref{eqB2})
and (\ref{eqC1})-(\ref{eqD2}) defined in vacuum, are unmodified
in the medium, since the pion mass in vacuum is used.

Our estimates of the in-medium couplings, $g_{\pi B B}^*$,
relative to those in vacuum $g_{\pi B B}$~\cite{CBMTony,TsushimaCBM},
rely on the Goldberger-Treimann relation~\cite{GTrelation}.
The in-medium to the free coupling constant ratio may be expressed by,
\ba
\frac{g_{\pi BB}^*}{g_{\pi BB}}
&=& \left(\frac{f_\pi}{f_\pi^*}\right)
\left(\frac{g_A^{B*}}{g_A^B}\right) \left(\frac{M_B^*}{M_B}\right),
\nonumber\\
&\simeq& \left(\frac{f_\pi}{f_\pi^*}\right)
\left(\frac{g_A^{N*}}{g_A^N}\right) \left(\frac{M_B^*}{M_B}\right),
\label{EqGTrelation}
\ea
where $f_\pi$, $g_A^B$ and $M_B$ are respectively the
pion decay constant, axial coupling constant of the baryon $B$ and
its mass, and the corresponding quantities in nuclear matter with $^*$.
First for $f_\pi^*$, it was estimated in  \cite{fpiMedium}
and we use the $f_\pi^{(t)}$ in nuclear matter as $f_\pi^*$ above.
On the other hand, $g_A^{N*}/g_A^N$ was estimated in QMC~\cite{QMCgA},
and $M_B^*/M_B$
as in table \ref{tableMass}.
For $g_{\pi\Lambda\Sigma}^*$, the relevant quantities
different from the diagonal cases are $M_\Lambda^*/M_\Lambda$
or $M_\Sigma^*/M_\Sigma$.
To estimate the maximally modified case, we use
$M_\Lambda^*/M_\Lambda$, although the difference is less
than 1\%. The modification of the coupling constants
for the densities $\rho=0.5 \, \rho_0$ and $\rho_0$
are summarized in table \ref{tableCoupling}.
In table \ref{tableCoupling} the in-medium coupling constants $g_{\pi BB'}^*$
via equation (\ref{EqGTrelation}) either decrease, or remain close
to the vacuum values. In QMC it is expected that the values slightly
decrease in medium, since the coupling is the same as that of the weak axial
coupling constant $g_A$, namely $\gamma^\mu\gamma_5$ to the Dirac spinor,
and $g_A^{N*}$ in nuclear matter decreases~\cite{QMCgA}.
Some unexpected behavior for
$g_{\pi\Sigma\Sigma}^*$ and $g_{\pi\Xi\Xi}^*$
are due to the large decrease of
$f_\pi^*$ estimated in  \cite{fpiMedium}.
The decreasing rate of $f_\pi^*$ in medium
overcomes slightly that of the $g_A^{N*}  M_{\Sigma}^*$
and $g_A^{N*} M_{\Xi}^*$,
and thus $g_{\pi\Sigma\Sigma}^*$ and $g_{\pi\Xi\Xi}^*$
increase slightly.
Thus, the trend of changing in values for
$g_{\pi BB'}^*$ in nuclear matter, is consistent with
that expected from QMC, although the latter generally
does not contain the pion cloud effects, and
it is this pion ($f_\pi^*$) which leads to slightly
unexpected density dependence for
$g_{\pi\Sigma\Sigma}^*$ and $g_{\pi\Xi\Xi}^*$.

\begin{table}[t]
\begin{center}
\begin{tabular}{l c c c}
\hline
\hline
      &$\rho=0$ &$\rho=0.5 \,\rho_0$ &$\rho=\rho_0$ \\
\hline
$g_{\pi NN}^*/g_{\pi NN}$                      &1  &0.921  &0.899 \\
$g_{\pi\Lambda\Sigma}^*/g_{\pi\Lambda\Sigma}$  &1  &0.973  &0.996 \\
$g_{\pi\Sigma\Sigma}^*/g_{\pi\Sigma\Sigma}$    &1  &0.977  &1.004 \\
$g_{\pi\Xi\Xi}^*/g_{\pi\Xi\Xi}$    &1  &1.012  &1.067 \\
\hline
\hline
\end{tabular}
\end{center}
\caption{
Modification of the $\pi B B'$ coupling constants in
nuclear matter with $\rho_0=0.15$ fm$^{-3}$.
}
\label{tableCoupling}
\end{table}

Next, we comment on the effects of the baryon mass modifications in the
intermediate baryon propagator in the nuclear medium.
In the pion cloud dressing of the current
shown in figure \ref{figPionCloud} in vacuum,
the intermediate states baryons belong to the same isomultiplet
except for the $\Lambda$ and $\Sigma$ cases.
Therefore the intermediate baryon masses are the same
or at most differ 1\% ($\Lambda$ and $\Sigma$).
In-medium mass modifications of the baryons $B'$
apply in the same way as those for the initial and final baryons,
and thus the mass difference between the baryons $B$ and $B'$
in the nuclear medium is the same as those in vacuum, namely,
$[M_{B'} - M_B] \simeq [M_{B'}^* - M_B^*]$,
and we can approximately have the same pion cloud effects
due to the modifications of the baryon masses.
Thus, the modification of the pion cloud effects
in the nuclear medium arises entirely  from the
modification of the pion-baryon coupling constants,
$g_{\pi BB'}^*$, in the present approach.

Finally, we estimate the in-medium modifications
for the pion cloud dressing
by replacing the constants those in vacuum
$\beta_B$ ($B=N,\Lambda,\Sigma, \Xi$)
given by  equations (\ref{EqbetaL})-(\ref{EqbetaX}),
by the new constants $\beta_B^\ast$
to be presented next.
We keep the value $\alpha=0.6$, the same as that in the vacuum,
and use the in-medium
coupling constants $g_{\pi BB'}^*$, and get
$\beta_B^*$ for $B=N,\Lambda,\Sigma,\Xi$,
\ba
\beta_N &\to&
\beta_N^* =  \left( \frac{g_{\pi NN}^*}{g_{\pi NN}} \right)^2,
\label{WqbetaNs}\\
\beta_\Lambda &\to&
\beta_\Lambda^* = \frac{4}{3} \alpha^2
\left( \frac{g_{\pi\Lambda\Sigma}^*}{g_{\pi\Lambda\Sigma}} \right)^2,
\label{EqbetaLs} \\
\beta_\Sigma &\to&
\beta_\Sigma^* = 4(1-\alpha)^2
\left( \frac{g_{\pi\Sigma\Sigma}^*}{g_{\pi\Sigma\Sigma}} \right)^2,
\label{EqbetaSs} \\
\beta_\Xi &\to&
\beta_\Xi^* = (1-2 \alpha)^2
\left( \frac{g_{\pi\Xi\Xi}^*}{g_{\pi\Xi\Xi}} \right)^2.
\label{EqbetaXs}
\ea
Note that according to the relation between
the constants $\beta_B$ and the normalization constants $Z_B$
[see equations (\ref{eqZB})]
the renormalization
constants will also be modified to $Z_B^\ast$
(replacing $\beta_B \to \beta_B^\ast$).

\section{Results}
\label{secResults}

According to the new constraints imposed
on the model (chiral symmetry and fit to the experimental square radii)
we readjust the parameters of the model
and carry out a new calibration
that differs from the one presented
in  \cite{OctetEMFF}.

We divide the presentation of our results in two parts.
First, we will present the results of the new calibration
for the vacuum, mainly concentrating on the nucleon form factors.
Next, we will present the results
for the baryon octet electromagnetic form factors
in the nuclear medium,
showing also each contribution
from the valence quarks and pion cloud.

\subsection{Octet baryon electromagnetic form factors in vacuum}

We present here the results
of the global fit in vacuum, and compare with
the previous model in  \cite{OctetEMFF}.
The values of the valence quark parameters
including the momentum-range scales of the
octet baryon wavefunctions and
the $u$ ($\kappa_u$) and $d$ ($\kappa_d$)
quark anomalous magnetic moments are
presented in table \ref{tabBare}.
As for the parameters associated with
the pion cloud, they are given in table \ref{tabPionCloud}.

We fit the parameters of the model
to the  nucleon data~\cite{JlabR1,Puckett10,Arrington07,MainzR1,Passchier99,Eden94,JlabR2,Madey03,Riordan10,Schiavilla01,Bosted95,MainzR2,Lachniet09,Zhan11},
the octet lattice QCD data,
the octet magnetic moments ($\Lambda, \Sigma^{+,-}, \Xi^{0,-}$)~\cite{PDG},
the nucleon electric and magnetic radii
and also the $\Sigma^-$ electric radius.
Following  \cite{OctetEMFF} we use
some constraints in the fit to the
octet lattice and physical data.
The lattice data considered are the data
with $m_\pi= 351, 591$ MeV from  \cite{Lin09}
for $n,p,\Sigma^\pm,\Xi^{0,-}$.
For these pion mass values, the pion cloud effects are
expected to be small.

To perform a fit we have taken care of the following points
in order to achieve a fair description of {\it both}
the physical and lattice data:
\begin{enumerate}
\item
The impact from the nucleon physical form factor data
is doubled
compared with the octet lattice form factor data.
\item
Statistical errors are doubled for the neutral particles $n$ and $\Xi^0$
to take into account the possible systematic errors in the lattice.
\item
We double the error bars for the physical magnetic moment data
to avoid dominating $\chi^2$ due to the extremely accurate measurements
in comparison with the nucleon physical form factor data.
\item
The impact from the nucleon radii data are doubled to be consistent with
the nucleon form factor data.
\item
The experimental error for the $\Sigma^-$ electric square radius
is reduced in our fit
(from 0.15 fm$^2$  to 0.015 fm$^2$)
in order to avoid the dominance of the
nucleon square radii data in the fit and also
to get the same order of contribution from
the pion cloud as that for the proton.
\end{enumerate}
Below we explain in detail the fitting procedure described above.

The first condition, the reinforcement of the impact
of physical data in the fit is included since the number
of physical data points (202) is inferior to
the number of lattice data points (272)
as discussed in  \cite{OctetEMFF}.
The reduction of the impact of the
neutral particle lattice data was also discussed in detail
in  \cite{OctetEMFF}.
As for the magnetic moment data
(magnetic form factors for $Q^2=0$)
the fitting condition is included to avoid the excessive
dominance of the $Q^2=0$ data.
The first 3 conditions follow the procedure
used in the previous work
where the nucleon physical data
and octet magnetic moments and the
octet lattice data were used to
calibrate the model~\cite{OctetEMFF}.
We discuss now the inclusion of the
data related with the electric and magnetic radii
that explain the last two conditions.

As discussed, the available experimental information
about the octet radii is restricted to
the nucleon ($p$ and $n$) electric and magnetic radii
and the $\Sigma^-$ electric radius.
The nucleon square radii are given in table \ref{tabRadiiN}.
For the $\Sigma^-$ electric square radius
the experimental value is
$r^2_{E\Sigma^-}=0.61\pm0.15$ fm$^2$~\cite{Eschrich01}.
In the $\chi^2$ calculation we use
a factor 2 for the radii data
to be consistent with the
physical and lattice data in the $\chi^2$ evaluation
(impact 2, the same as the nucleon form factor data).
The inclusion of the octet radii data is important
to calibrate the low $Q^2$ regime of the model.
In addition, the radii data constrain the degrees
of freedom related to the pion cloud dressing,
and reflect the effect of the chiral symmetry.
Note however the large error associated
with the $\Sigma^-$ electric square radius.
Although the nucleon radii data
restrict the possible values to a small interval,
the $\Sigma^-$ electric square radius has a large
interval of variation.

The minimization of the $\chi^2$
using the result for the $\Sigma^-$ electric square radius
(with the error 0.15 fm$^2$)
leads to a very large $\Sigma^-$ electric square radius
compared with that of the proton,
as a consequence of much larger contribution
from the pion cloud than that for the proton.
A large pion cloud contribution for
the $\Sigma^-$ electric square radius contradicts
what it is expected from $SU(3)$ symmetry,
where the contribution is expected to be of comparable
with that for the proton.
Also, estimates of the pion cloud contribution
from  \cite{Wang09} for the electric square radii
predict similar contributions for
the proton and $\Sigma^-$.
In order to impose a small pion cloud contribution
to the $\Sigma^-$ electric square radii with the amount
close the the proton case, we reduce the
error of $\Sigma^-$ electric square radius in the $\chi^2$
calculation to 0.015 fm$^2$ (instead of 0.15 fm$^2$).
We consider then,
for the $\Sigma^-$ electric square radius an error bar
comparable with the error bars from the nucleon radii.
With the choice 0.015 fm$^2$,
the contribution of the $\Sigma^-$ electric radius
to $\chi^2$ is of the same order
as the other contributions
(particularly the proton magnetic square radius),
and the calibration of the model
is not dominated by a particular observable.
An additional motivation to use
a standard deviation for the $\Sigma^-$
electric square radius with a magnitude
$\approx 0.02$ fm$^2$ is that we can
achieve a simultaneous good description
of the nucleon form factor data
($\chi^2$ per data point $\approx 2$)
and lattice data ($\chi^2$ per data point $\approx 6$),
as in the previous work,
where the experimental information about the
baryon radii was not taken into consideration~\cite{OctetEMFF}.
We note that this is an {\it ad hoc} procedure,
but in the absence of additional information
about the quark core effect (or pion cloud)
in the physical regime, it is the simplest
way to make a realistic calibration
of the present model.
Later we will discuss the sensitivity of the final
fit to the values of the $\Sigma^\pm$ electric radii.

\begin{table}[t]
\begin{center}
\begin{tabular}{c c c c}
\hline
\hline
      $\beta_1$ & $\beta_2$ & $\beta_3$  & $\beta_4$  \\
\hline
      0.0532 & 0.809   &  0.603   &   0.381 \\
\hline
\hline
                &  $\kappa_u$ & $\kappa_d$ & $\kappa_s$ \\
\hline
                &   1.711    & 1.987    & 1.462 \\
\hline
\hline
\end{tabular}
\end{center}
\caption{
Parameters associated with the valence quarks
determined by the fit.
The value of $\kappa_s$ was determined in  \cite{Omega}
in the study of the baryon decuplet.
The analytical form of the scalar wavefunctions,
depending on $\beta_i$ ($i=1,..,4$) are
presented in section \ref{secBare}.
}
\label{tabBare}
\end{table}

From the parameters associated
with the valence quark contributions in table \ref{secBare},
we note that
$\beta_1 < \beta_2,\beta_3, \beta_4$
as expected from the interpretation of $\beta_1$,
as the 3-quark long range parameter
according to equations (\ref{eqPsiSN})-(\ref{eqPsiSXi}).
Also the order of the momentum range scales
$\beta_2 > \beta_3 > \beta_4$
are consistent with the fact
that the system with only light quarks
(parameter $\beta_2$) is more spread
than a system with one strange quark (parameter $\beta_3$),
and that it is less compact than
a system with two strange quarks (parameter $\beta_4$)
in the position space.
The same trend was {\it observed} in  \cite{OctetEMFF}.

As for the values of the pion cloud parameters
the more significant difference is the increase
of $B_1$, which implies an increase
of the pion cloud contribution for $G_E$,
for the all members of the octet baryons, near $Q^2=0$.
Taking the nucleon case as an example,
$Z_N=1/(1+3 B_1) \simeq 0.87$ in the vacuum,
means that the pion cloud contribution for
the proton charge is about 13\%.
This is still rather small compared with a model
such as CBM, yet in comparison
with the previous work,
the low $Q^2$ behavior of
the functions $\tilde B_1$ and $\tilde B_2$
is modified. (See section \ref{secPionCloud2}).
This modification changes significantly the
behavior of the pion cloud parametrization.
Nevertheless, the new parameter values are close
to the previous ones except for $C_2$ \cite{OctetEMFF},
which is now much smaller
(reduction in the effect of the Dirac
form factors due to the process
shown in figure \ref{figPionCloud}(b) to the magnetic form factors).

\begin{table}[t]
\begin{center}
\begin{tabular}{c c c c c }
\hline
\hline
    \sp $B_1,B_2$ \sp \sp & \sp\sp\sp $C_2$ \sp\sp\sp
                          & \sp $D_1^\prime,D_2$ \sp &
\sp $b_1^\prime, b_2^\prime$ \sp
    & \sp $\Lambda_1,\Lambda_2$(GeV)  \\
\hline
      0.0510  &             &     -0.148    & \sp 1.036 &  0.786   \\
      0.216  &  0.00286    & \sp  0.0821    &    -1.987 &  1.132   \\
\hline
\hline
\end{tabular}
\end{center}
\caption{
Parameters associated with the pion cloud dressing
determined by the fit.
See the parametrization of the functions
$\tilde B_i, \tilde C_i$ and $\tilde D_i$ ($i=1,2$)
in section \ref{secPionCloud}.
The values of $b_1^\prime,b_2^\prime$ are
calculated from the equations (\ref{eqR1b})-(\ref{eqR2b}).
}
\label{tabPionCloud}
\end{table}

Comparing the quality of the present fit
with the previous one, we have obtained
a slightly worse description of the
lattice QCD data
[$\chi^2$ per data point of 6.0
to be compared with 5.0],
and also a less accurate description
of the nucleon data
[$\chi^2$ per data point of 1.99
to be compared with 1.93].
In detail we have now per data point,
\ba
& &\chi^2(G_{Ep})=1.89, \hspace{.3cm} \chi^2(G_{Mp})=1.69,
\nonumber \\
& &\chi^2(G_{En})=1.78, \hspace{.3cm} \chi^2(G_{Mn})=2.41,
\ea
to be compared with the previous values of
$\chi^2(G_{Ep})=1.60$, $\chi^2(G_{Mp})=1.87$,
$\chi^2(G_{En})=1.86$ and $\chi^2(G_{Mn})=2.27$ \cite{OctetEMFF}.
In simple words we have improved the description
of $G_{En}$ but we have lost some precision
in the description of the other form factors,
particularly for $G_{Mn}$.
The neutron magnetic form factor is difficult
to describe as a consequence of the high accuracy of
recent data, that disagree with the previous sets
(see details in  \cite{OctetEMFF}).
The loss of precision in some
nucleon physical data is compensated
by the quality of the description
of the nucleon and $\Sigma^-$ square radii,
where the constraints
from chiral symmetry are imposed.

The main electromagnetic properties of the
octet baryons in vacuum, such
as magnetic moments (in $\hat \mu_N$ units),
the electric and magnetic square radii
are presented in Table~\ref{tabOctet}
and compared with the experimental results.

The square radii are defined according to
\ba
& &
r_{EB}^2= - \left. \frac{6}{G_{EB}(0)}\frac{d G_{EB}}{d Q^2} \right|_{Q^2=0},
\label{eqREX}
\\
& &
r_{MB}^2= - \frac{6}{G_{MB}(0)}\left. \frac{d G_{MB}}{d Q^2} \right|_{Q^2=0}.
\label{eqRMX}
\ea
Note that the square radii are normalized by
the value of the form factor ($G_{EB}$ or $G_{MB}$)
at $Q^2=0$, the usual definition.
For neutral particles
[$G_{EB}(0) =0$], we use the same definition with
$G_{EB}(0) \to 1$.

\begin{table*}[t]
\begin{center}
\begin{tabular}{l r r r r r r r}
\hline
\hline
  $B$   & $\mu_B$ & $\mu_B^{exp}$  & &$r_{EB}^2$ & $(r_{EB}^2)_{exp}$
& $r_{MB}^2$ & $(r_{MB}^2)_{exp}$ \\
\hline
$p$ &  2.737  &  2.793  & &  0.782 & 0.763(14) & \sp 0.718 & \sp \sp0.698(11)  \\
$n$ & -1.933 &  $-$1.913 & & -0.113 & $-$0.1161(22)& 0.729 & 0.743(13) \\
  &        &         &     &     &     &      \\
$\Lambda$  & -0.628 & -0.613(4)  & & 0.068    &     & 0.228    &      \\
 &        &         & &    &     &     &      \\
$\Sigma^+$ &  2.600      &  2.45(2)   & & 0.713  &     & 0.516    &      \\
$\Sigma^0$ &  0.728      &            & & 0.039  &    &  0.388   &      \\
$\Sigma^-$ & -1.143     &  $-$1.16(3)    & & 0.643 & 0.61(15)    & 0.642    &      \\
           &            &         &     &     &  &   &      \\
$\Xi^0$ & -1.488      &  $-$1.250(14)  & & 0.097  &     &  0.319   &      \\
$\Xi^-$ & -0.689      &  $-$0.65(3)    & & 0.403  &     &  0.268   &      \\
\hline
\hline
\end{tabular}
\end{center}
\caption{Electromagnetic proprieties of
octet baryons in vacuum.
See Table~\ref{tabRadiiN} for the description of
the nucleon radii data.
Additional data are from  \cite{Eschrich01} ($\Sigma^-$ electric square radius)
and  \cite{PDG} (magnetic moments).  }
\label{tabOctet}
\end{table*}

In table~\ref{tabRadii} we present also the decomposition
of the octet baryon square radii into
valence or bare core ($b$) and  pion cloud ($\pi$) contributions.
These are defined for $X=E,M$, by
\ba
& &
(r_{XB}^2)_b =
 - \left.
Z_B \frac{6} {G_{XB}(0)} \frac{d G_{X0B}}{d Q^2} \right|_{Q^2=0},
\label{eqRXbare}
\\
& &
(r_{XB}^2)_\pi =
 - Z_B \frac{6}{G_{XB}(0)}\left. \frac{d (\delta G_{XB})}{d Q^2}
\right|_{Q^2=0},
\label{eqRXpion}
\ea
based on the decomposition~(\ref{eqGX1}).
Note that the components $(r_{XB}^2)_b$ and $(r_{XB}^2)_\pi$
are normalized by the total form factor $G_{XB}(0)$.

In the fitting process we can conclude that the model
is very sensitive to the neutron data.
This can be a consequence of the impact
of the pion cloud in the neutron form factors.
[This can also be true for other neutral
particles, but there is much more information about the neutron.]
A very useful index to study the
neutron form factors is the electric square radius.
The neutron electric square radius quantifies the slope
of the $G_{En}$ form factor at $Q^2=0$.

In the present model we can quantify the
contribution of the valence quarks
and the pion cloud using the
decomposition (\ref{eqRXbare}) and~(\ref{eqRXpion})
for the neutron electric radius ($X=E$).
In the present case the result is
\ba
r_{En}^2= (-0.097)_b+ (-0.016)_{\pi} \;\; \mbox{fm}^2.
\label{eqREn}
\ea
From this we conclude that the pion cloud gives
about 14\% of the total result [$-0.113$ fm$^2$,
very close to the experimental result $-0.116$ fm$^2$].
Again the pion cloud contribution is still much smaller
than suggested by an explicit calculation in
the CBM~\cite{CBMTony}.
Nevertheless, this is the best we have been able
within the constraints of the fit to the whole octet.
We note that the sign of the
bare contribution is
consistent with the lattice QCD data as we will explain next.
Since $G_{En}$ is positive and increases
with $Q^2$ (positive derivative)
it gives a negative contribution for $r_{En}^2$,
according to the definition~(\ref{eqREX})
[with the replacement $G_{En}(0) \to 1$].

\begin{table*}[t]
\begin{center}
\begin{tabular}{l r r r r r r}
\hline
\hline
  $B$   & $(r_{EB}^2)_b$ &  $(r_{EB}^2)_\pi$ &
$r_{EB}^2$ & $(r_{MB}^2)_b$ & $(r_{MB}^2)_\pi$ & $r_{MB}^2$ \\
\hline
$p$ &  0.614 & 0.168 & \sp  0.782 & 0.601 & 0.117 & \sp 0.718  \\
$n$ & -0.097 &-0.016 & -0.113 & 0.624 & 0.105 & 0.729 \\
    &        &       &        &     &     &      \\
$\Lambda$  & -0.005 &  0.073  & 0.068 & 0.449 & -0.221 & 0.228  \\
    &        &       &        &     &     &      \\
$\Sigma^+$  & 0.470 &  0.244  & 0.713 & 0.350 & 0.166 & 0.516  \\
$\Sigma^0$  & -0.001 & 0.040  & 0.039 & 0.291 & 0.097 & 0.388  \\
$\Sigma^-$  &  0.480 & 0.162  & 0.643 & 0.388 & 0.253 & 0.642  \\
    &        &       &        &     &     &      \\
$\Xi^0$  & 0.096 &  0.001  & 0.097 & 0.325 &  -0.005 & 0.319  \\
$\Xi^-$  & 0.382 &  0.021  & 0.403 & 0.218 &  0.050 & 0.218  \\
\hline
\hline
\end{tabular}
\end{center}
\caption{Octet baryon square radii
in vacuum, and the bare and pion cloud contributions.}
\label{tabRadii}
\end{table*}

The effects of the
valence quarks and pion cloud effects for
the octet baryon radii were also estimated in  \cite{Wang09}.
There, the formalism used is the
heavy baryon chiral perturbation theory
applied to the lattice QCD data from  \cite{Boinepalli06}.
In that work of Wang {\it et al.}~\cite{Wang09}
the quenched lattice results are
corrected by subtracting the quenched
and finite volume effects to extract
the full QCD result in the physical limit of infinite volume.
In this case we estimate the bare contribution
from the quenched QCD contribution (QQCD),
since in that work it is the component that
better approximates the valence quark content.
For simplicity we will ignore the uncertainties
of the estimates.
In that case pion cloud is the
dominant contribution (67\%) and
the slope of the core contribution is positive.
The last result suggests that the contribution
of the quark core to $G_{En}$ is positive.
This is qualitatively consistent with our results,
but it is important to quantify
the contribution of each term.

\begin{figure*}[t]
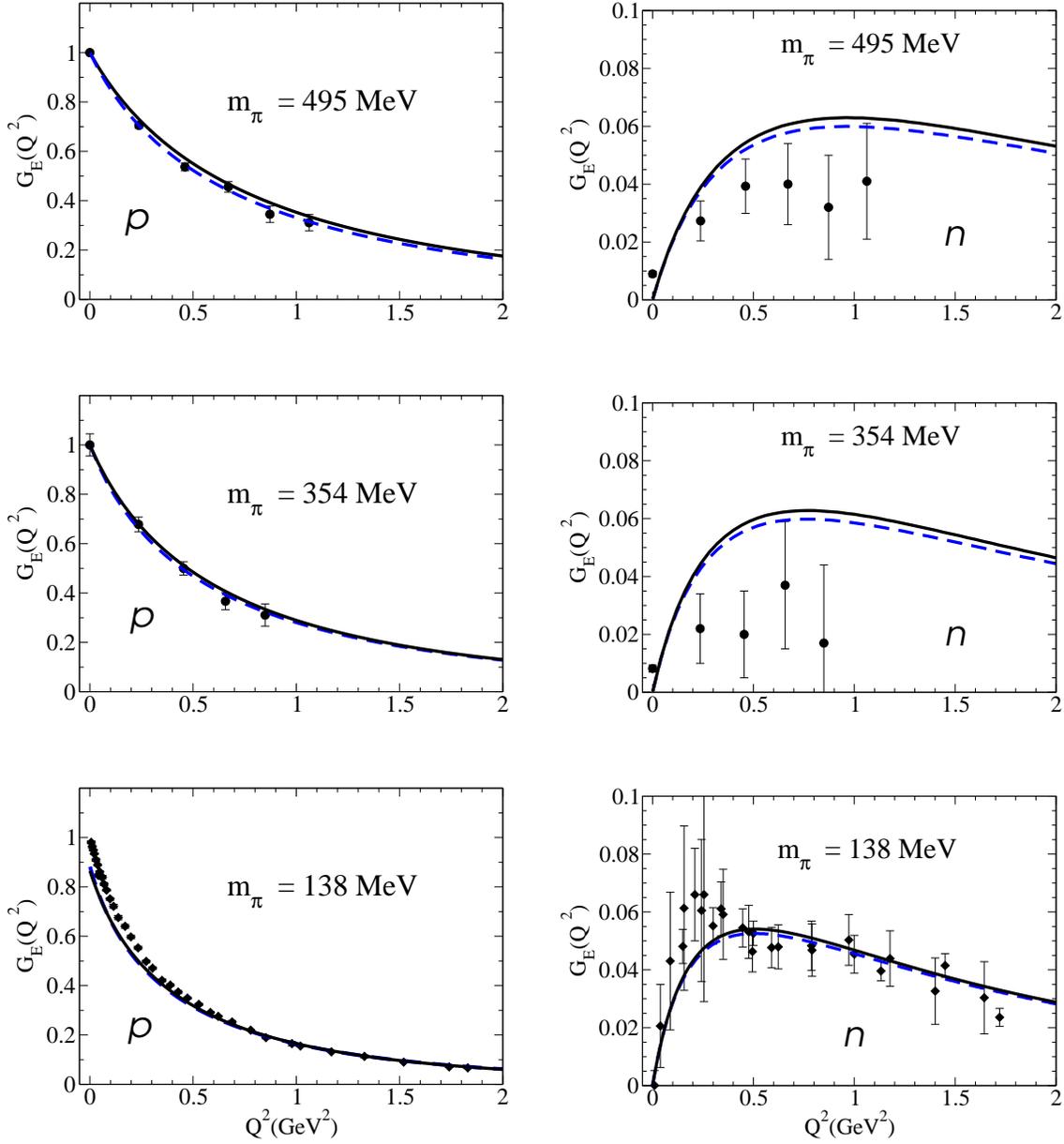

\centerline{\vspace{0.4cm}  }
\centerline{
\mbox{
\includegraphics[width=2.8in]{GEp495_1.eps}   \hspace{.5cm}
\includegraphics[width=2.8in]{GEn495_1.eps}
 }}
\vspace{1.cm}
\centerline{
\mbox{
\includegraphics[width=2.8in]{GEp354_1.eps}   \hspace{.5cm}
\includegraphics[width=2.8in]{GEn354_1.eps}  }}
\vspace{1.cm}
\centerline{
\mbox{
\includegraphics[width=2.8in]{GEp138_1.eps}   \hspace{.5cm}
\includegraphics[width=2.8in]{GEn138_1.eps}  }}
\caption{\footnotesize{
Nucleon electric form factors calculated in lattice,
the present model (solid line), and the model from
 \cite{OctetEMFF} (dashed line).
For the physical point $m_\pi=138$ MeV, only the contributions
from the core are included (without the pion cloud).
}}
\label{figNucleon}
\end{figure*}

\begin{figure*}[t]
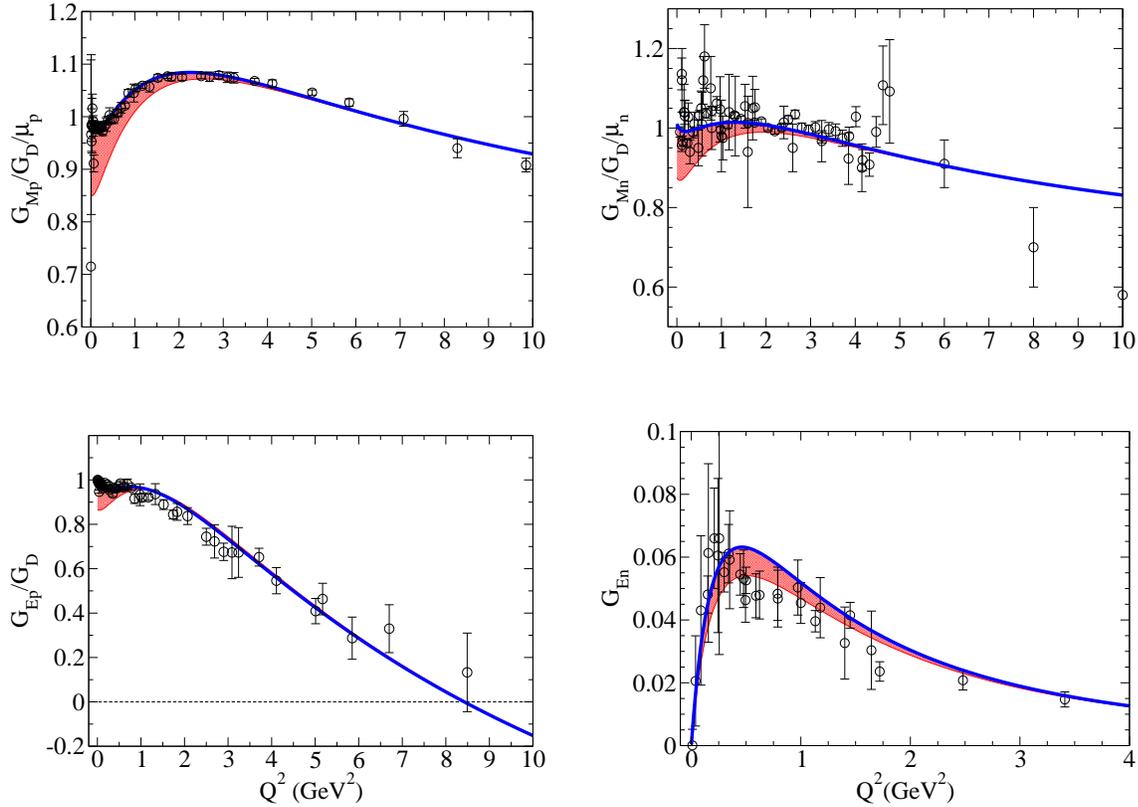

\centerline{\vspace{0.4cm}  }
\centerline{
\mbox{
\includegraphics[width=2.8in]{GMp138_3.eps}   \hspace{.5cm}
\includegraphics[width=2.8in]{GMn138_3.eps}  }}
\centerline{\vspace{.5cm}}
\centerline{
\mbox{
\vspace{1cm}
\includegraphics[width=2.8in]{GEp138_3.eps}   \hspace{.5cm}
\includegraphics[width=2.8in]{GEn138_3.eps}  }}
\caption{\footnotesize{
Nucleon form factors fitted at the physical point
compared with the experimental data.
The bands represent the pion cloud contributions.
}}
\label{figNucleon2}
\end{figure*}

In our model the valence quark contribution
is determined by the fit to the lattice QCD data
and also the nucleon physical data.
In both cases the neutron data are included.
As discussed in the previous work~\cite{OctetEMFF}
the spectator quark model
simulates the magnitude and sign of the
lattice QCD data~\cite{Lin09} for $G_{En}$
used in the calibration.
That is a consequence of the VMD parametrization
where there is an asymmetry between the
up ($u$) and down ($d$) quarks electromagnetic structure.
This is represented by a distinct parametrization
of the isoscalar $f_{1+}$ and isovector $f_{1-}$ quark form factors.
In particular we have
$f_{1+}(Q^2)> f_{1-}(Q^2)$, leading to the result\footnote{Using the
expression for the nucleon $G_E$
given by equation (28) of  \cite{Nucleon},
which is equivalent to the one used in this
work without pion cloud,
we have
\ba
G_{En}= \frac{1}{2}B \left[
(f_{1+}-f_{1-}) -\sfrac{Q^2}{4M^2} (f_{2+}- f_{2-})
\right],
\nonumber
\ea
where $B$ is the overlap integral (normalized to 1 at $Q^2=0$).
As the Pauli form factors are suppressed in the small $Q^2$ region,
the difference between $f_{1+}$ and $f_{1-}$ determines
the sign of $G_{En}$ near $Q^2=0$.
Taking $G_{Ep}-G_{En}$ we get a dependence in the isovector
form factors
$f_{i-}$ ($i=1,2$).}
$G_{En}(Q^2)> 0$ for lattice and also for
the bare core in the physical case.
We note that although
quenched QCD simulations preserve
the isospin symmetry,
the full QCD simulations and the nature
violate (in small degree) the isospin symmetry.

In figure \ref{figNucleon} we present the results
of the model for the
nucleon electric form factors in lattice ($G_E$),
compared with the lattice data
from  \cite{Lin09} used in the present fit
for the cases $m_\pi=354, 495$ MeV.
The calculation is made using the
model extended to the lattice regime
based on the VMD parametrization
[see  \cite{OctetEMFF} for more details].
The contributions of the bare core
for the physical case ($m_\pi= 138$ MeV)
are also presented and compared with
the data.
In figure \ref{figNucleon} we present also the results of the previous
model~\cite{OctetEMFF}, which show
that the both models have similar results.

\begin{figure*}[t]
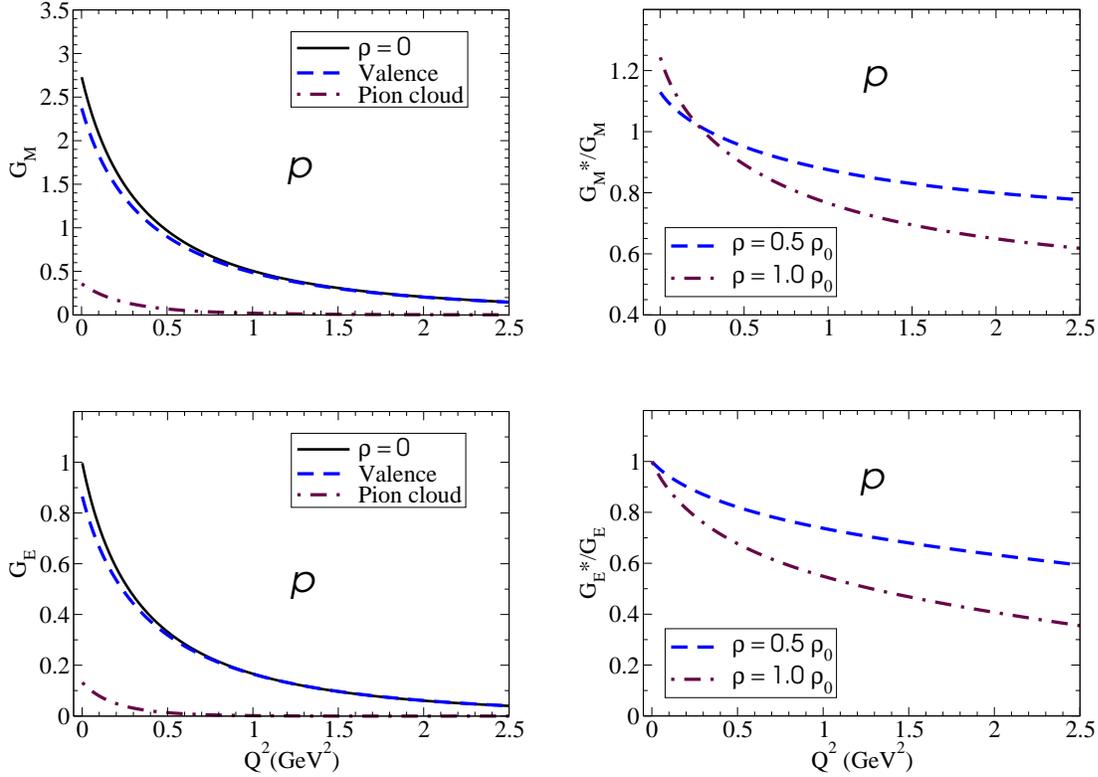

\centerline{\vspace{0.4cm}  }
\centerline{
\mbox{
\includegraphics[width=2.7in]{GMp.eps}   \hspace{.5cm}
\includegraphics[width=2.7in]{GMpR.eps}
}}
\centerline{
\vspace{0.5cm}  }
\centerline{
\mbox{
\includegraphics[width=2.7in]{GEp.eps}  \hspace{.5cm}
\includegraphics[width=2.7in]{GEpR.eps}
}}
\caption{\footnotesize{
Proton electromagnetic form factors calculated
for $\rho=0$ (vacuum) with a decomposition
of the valence and pion cloud contributions (left panel),
and the ratios to those of the $\rho=0$ (right panel)
for $\rho=0.5\, \rho_0$ (dashed line) and
$1.0 \, \rho_0$ (dash-doted line).
}}
\label{figProton}
\end{figure*}

\begin{figure*}[t]
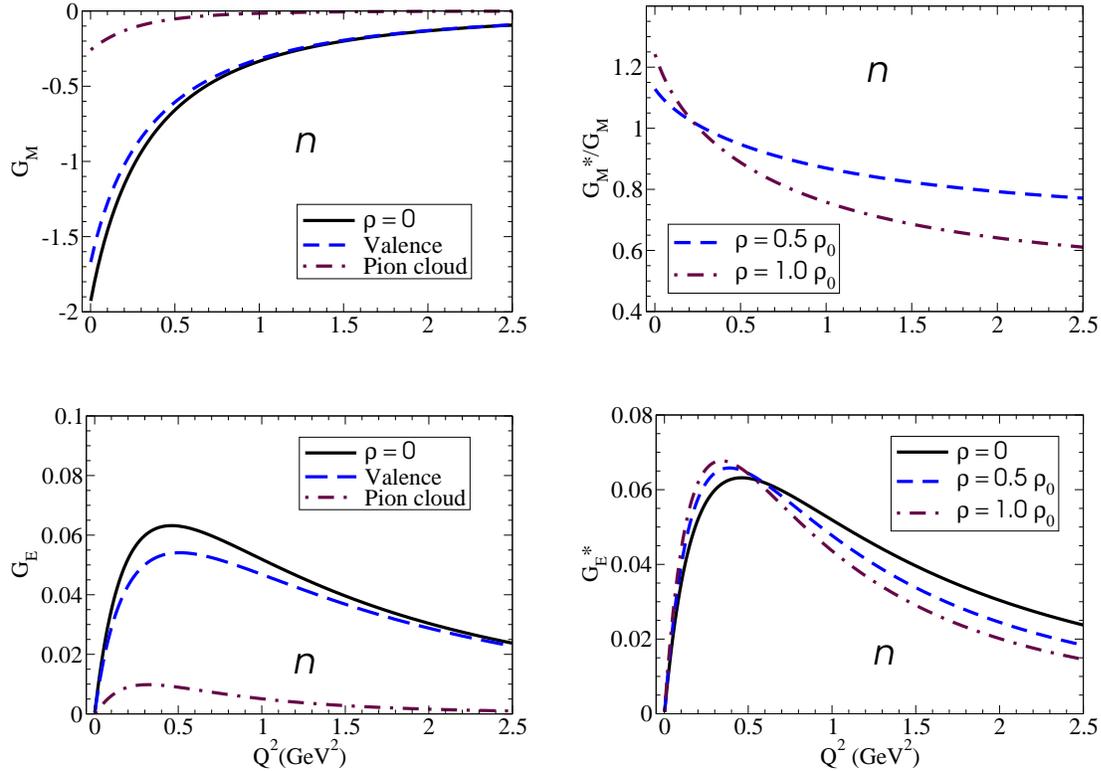

\centerline{\vspace{0.4cm}  }
\centerline{
\mbox{
\includegraphics[width=2.7in]{GMn.eps}   \hspace{.5cm}
\includegraphics[width=2.7in]{GMnR.eps}
}}
\centerline{
\vspace{0.5cm}  }
\centerline{
\mbox{
\includegraphics[width=2.7in]{GEn.eps}  \hspace{.5cm}
\includegraphics[width=2.7in]{GEnR.eps}
}}
\caption{\footnotesize{
Same as the caption of FIG.~\ref{figProton},
but for the neutron.
We note that, as discussed in the text,
some of the pion cloud contribution in models such as
CBM~\cite{CBMTony,TsushimaCBM} are included through
VMD in the valence terms in this parametrization.
}}
\label{figNeutron}
\end{figure*}

\begin{figure*}[t]
\centerline{\vspace{0.4cm}  }
\centerline{
\mbox{
\includegraphics[width=2.7in]{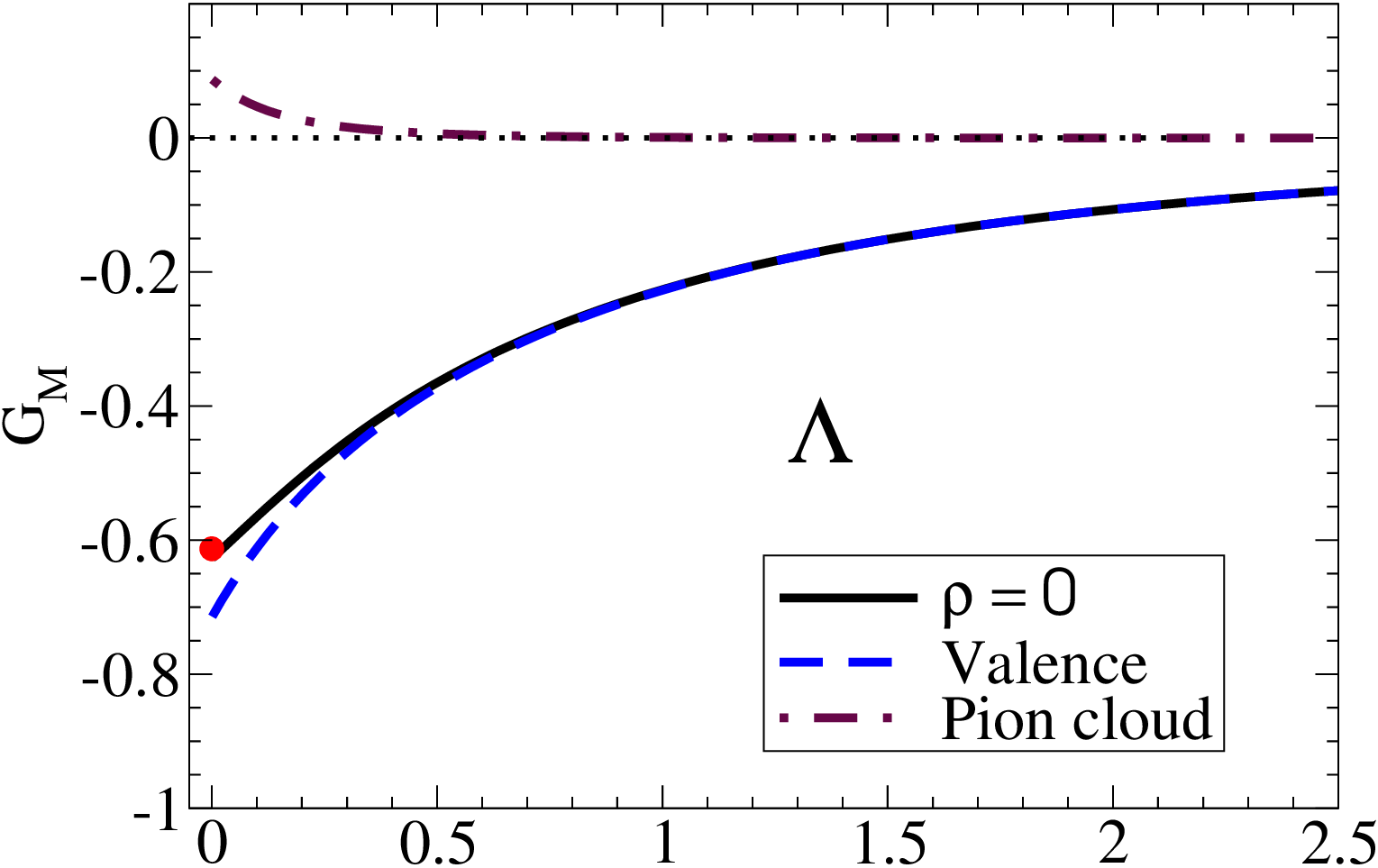}   \hspace{.5cm}
\includegraphics[width=2.7in]{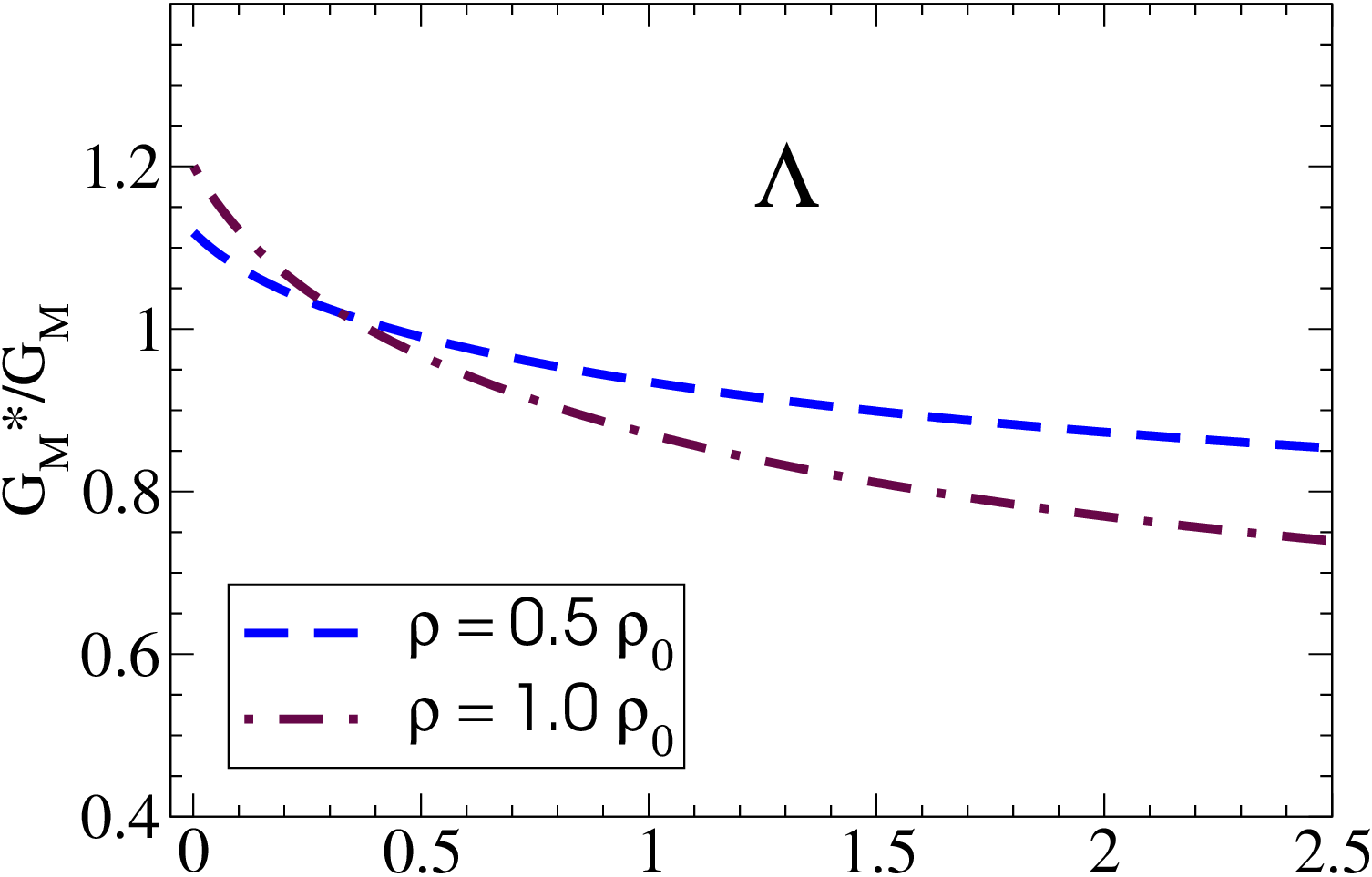}
}}
\centerline{
\vspace{0.5cm}  }
\centerline{
\mbox{
\includegraphics[width=2.7in]{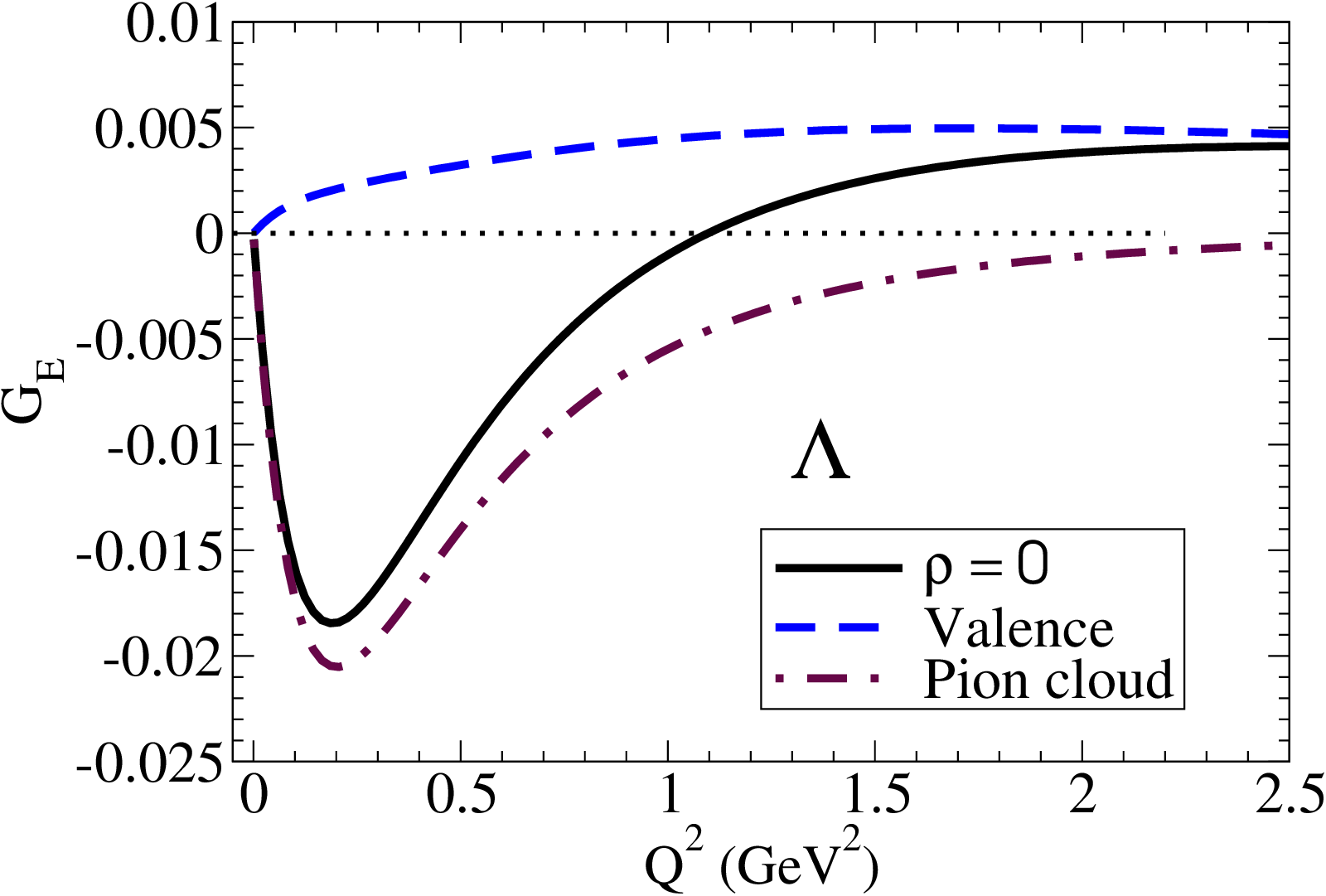}  \hspace{.5cm}
\includegraphics[width=2.7in]{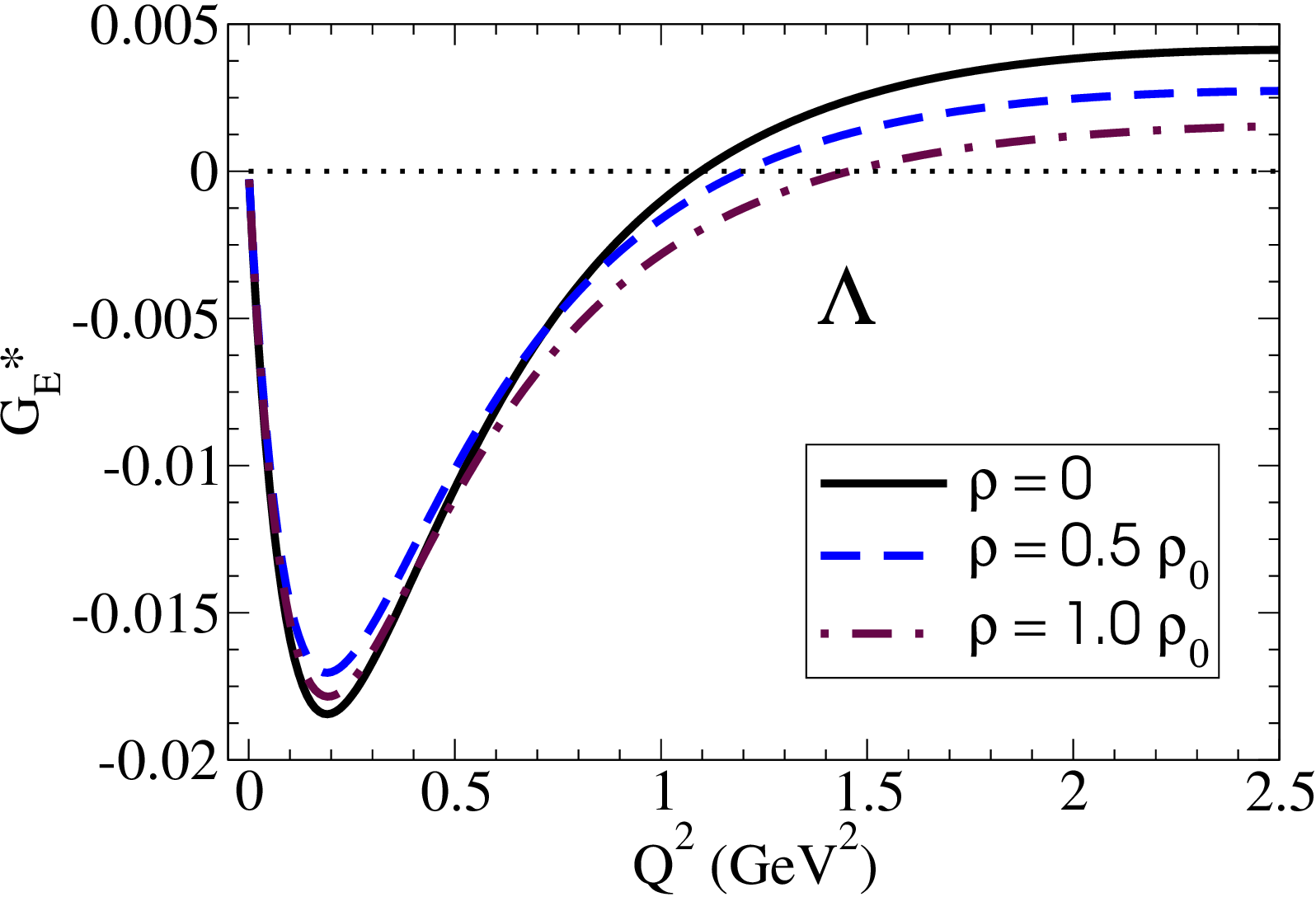}
}}
\caption{\footnotesize{
Same as the caption of FIG.~\ref{figProton},
but for $\Lambda$.
The experimental magnetic moment is shown by the (red) dot.
}}
\label{figLambda}
\end{figure*}

As we can observe from lattice data
($m_\pi=354$ MeV, and $m_\pi=495$ MeV)
for the neutron, the model results are close to the lattice data,
although there are some overestimates of the data.
Note also that the neutron lattice data are positive.
Since in this lattice QCD regime the pion cloud effects are
not important, we interpret the result as
a consequence of the isospin asymmetry
of the model for the valence quarks as already mentioned.
The figure supports our motivation
to apply the quark model to the lattice QCD regime,
using a model with isospin breaking.
One can argue however that
since we consider a quark current
parameterized by a VMD structure, some pion cloud effects
can also be in part of the current, particularly
in the term associated with the $\rho$-pole.
[For the high $Q^2$ region we can claim that the
quark structure is the only one that survives
consistently with the data.]
To analyze better this point we look again
for the neutron electric radius.
Since the $G_{En}$ data in lattice
show a positive increasing function of $Q^2$
in the low $Q^2$ region, this implies a negative contribution
of the bare core to the charge radius [see equation(\ref{eqREX})].
The simple comparison of our results
with those of Wang {\it et al.}~\cite{Wang09} based
on the relative contributions mentioned before
(67\% of pion cloud in $r_{En}^2$),
can lead a conclusion that about
53\% of our VMD parametrization
may be pion cloud
(67\% from Wang's result, and minus 14\%
in our estimate).
Notice, however, that Wang {\it et al.}~uses
a different structure for the pion cloud.
Contrary to a more common
representation for the pion cloud effects
in the nucleon system, as the one we
adopt (see figure \ref{figPionCloud}) following
 \cite{Miller02,Matevosyan05},
the work of Wang {\it et al.}~\cite{Wang09}
also includes a diagram with a pion double vertex
that comes from full QCD.
This diagram can also be a source
of the part for the $\sim 50\%$ difference
between our estimate and that of  \cite{Wang09}.
Future studies using lattice QCD simulation data
to reduce the model dependence for the
pion cloud contributions, combined with
a precise estimate for the valence quark
structure in the intermediate $Q^2$ region,
can help to pin down the effective contributions
of these degrees of freedom.

It can be useful for the nucleon case to compare
the lattice QCD data from  \cite{Lin09},
and our results using the other lattice QCD simulations.
There are for instance different data in  \cite{Lin10},
from the same group, than the data we have used in our calibration.
Unfortunately the lattice QCD studies
of the nucleon form factors are mainly performed
for the nucleon isovector form factors like
$G_E^V = G_{Ep}-G_{En}$ and $G_M^V = G_{Mp}-G_{Mn}$,
and results for $G_{En}$ have not been published in general.
Examples are the works of QCDSF collaboration~\cite{Gockeler05}
and the Cyprus group~\cite{Alexandrou06,Alexandrou11}.
The main reason to avoid determining the form factors for
proton and neutron separately,
is because the isoscalar form factors,
$G_E^S = G_{Ep}+ G_{En}$ and $G_M^S = G_{Mp}+ G_{Mn}$,
should include contributions from disconnected diagrams~\cite{Gockeler05},
which is a complex task with the present lattice QCD resources.
The results extracted from  \cite{Boinepalli06,Gockeler05,Collins11}
are too imprecise to draw any conclusion
about the magnitude and the sign of $G_{En}$ on lattice.
Nucleon isoscalar form factors were
calculated recently by the LHPC
collaboration~\cite{Syritsyn10}, and they show also a positive
result for $G_{En}$ for pion masses similar
to the ones in  \cite{Lin09},
but larger in magnitude.

Although there are other lattice simulations
of the nucleon form factors,
we keep our preference for the data in
 \cite{Lin09}, even if the data are affected
by some systematic errors as mentioned already~\cite{OctetEMFF}.
The main reason is the fact that
it is the only work extended to the octet baryons.
Therefore, the limitations in the neutron
form factor results can be compensated by the
inclusion of the data for $\Sigma^{+,-}$ and
$\Xi^{0,-}$.

The results of the fit for the nucleon
form factors are presented in figure \ref{figNucleon2}
and compared with the selected data~\cite{JlabR1,Puckett10,Arrington07,MainzR1,Passchier99,Eden94,JlabR2,Madey03,Riordan10,Schiavilla01,Bosted95,MainzR2,Lachniet09,Zhan11}
(see  \cite{OctetEMFF} for a detailed discussion
about the database).
In figure \ref{figNucleon2} the solid line gives the full result
(valence plus pion cloud) and the bands
the effect of the pion cloud.
The results for the remaining octet baryon members
will be presented in the next section
and extended to the nuclear medium.
Recall that except for the magnetic moments
there is no data for the $\Lambda$, $\Sigma$ and $\Xi$ systems.
Results in figure \ref{figNucleon2}
show the dominance of the quark core
and that the pion cloud effect is restricted
to the small $Q^2$ region.
As for the other octet members the
form factors are mainly determined by the lattice QCD data.
The fit to the lattice data
provides a good description (small $\chi^2$ per data point)
for the $\Sigma$ systems but are not so good
for the $\Xi$ systems.
As for the neutral particles,
$\Lambda, \Sigma^0$ and $\Xi^0$, our calibration
provide only a crude estimate of the
core effect, since these systems are not
constrained by lattice data (except for $\Xi^0$).
Therefore, the separation
between the core and pion cloud should be taken
with care particularly for $G_E$
(very small for neutral particles),
where small pion cloud effects and residual
quark core effects cannot be distinguished with precision.

\begin{figure*}[t]
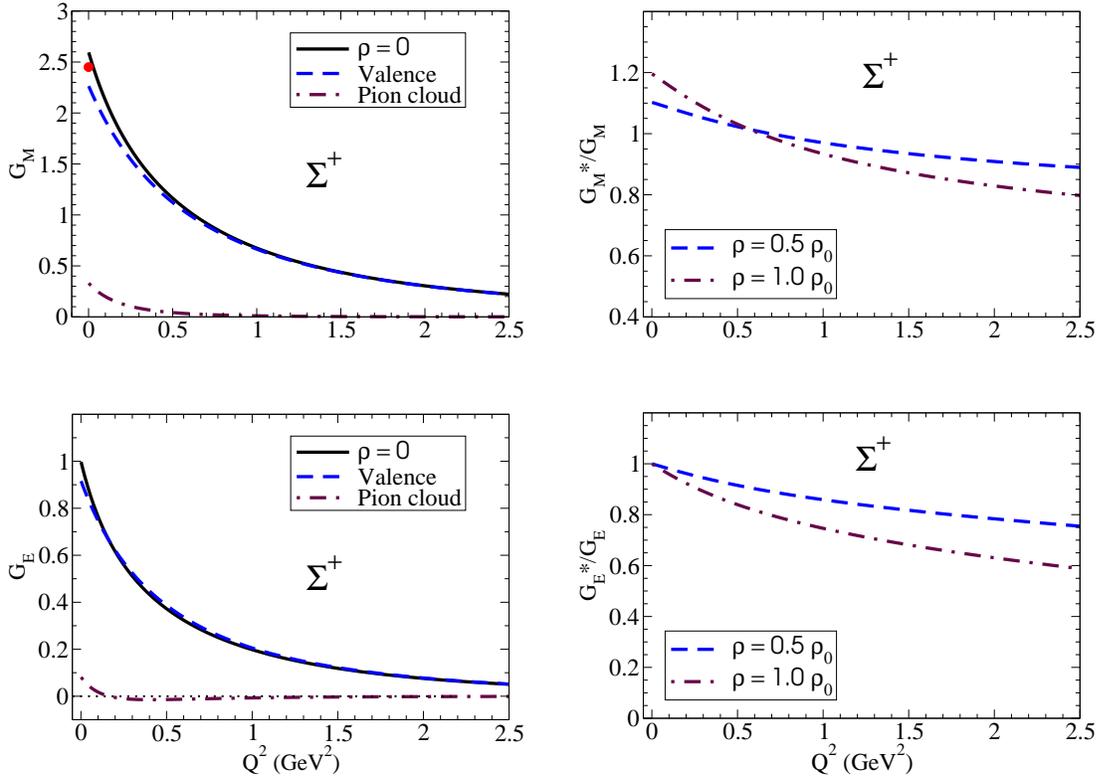

\centerline{\vspace{0.4cm}  }
\centerline{
\mbox{
\includegraphics[width=2.7in]{GMsigP.eps}   \hspace{.5cm}
\includegraphics[width=2.7in]{GMsigPR.eps}
}}
\centerline{
\vspace{0.5cm}  }
\centerline{
\mbox{
\includegraphics[width=2.7in]{GEsigP.eps}  \hspace{.5cm}
\includegraphics[width=2.7in]{GEsigPR.eps}
}}
\caption{\footnotesize{
Same as the caption of FIG.~\ref{figProton},
but for $\Sigma^+$.
The experimental magnetic moment is shown by the (red) dot.
}}
\label{figSigmaP}
\end{figure*}

Other  properties of the octet baryons in vacuum
are listed in Tables~\ref{tabOctet} and~\ref{tabRadii}.
From Table~\ref{tabOctet} we can conclude that
the present model can describe fairly well
the octet square radii
and the magnetic moments, except for $\Xi^0$.
From Table~\ref{tabRadii} we note
the dominance of the quark core effects on the
octet baryon radii.
The exception is the electric square radii
of the charge neutral particles, as mentioned already.
A note about
the large pion cloud contribution
for the $\Sigma^-$ magnetic square radii, is in order.
The large pion cloud contribution for $r_{M\Sigma^-}^2$
is essentially a consequence of the large contribution
of the term in $\tilde B_2$
(photon-pion coupling contribution for $F_{2B}$).
In fact the same effect appears for $\Sigma^+$,
but the final value is reduced by the larger
magnetic moment (factor 2.28),
in the definition of $r_{MB}^2$
[see equation (\ref{eqRMX})].

\begin{figure*}[t]
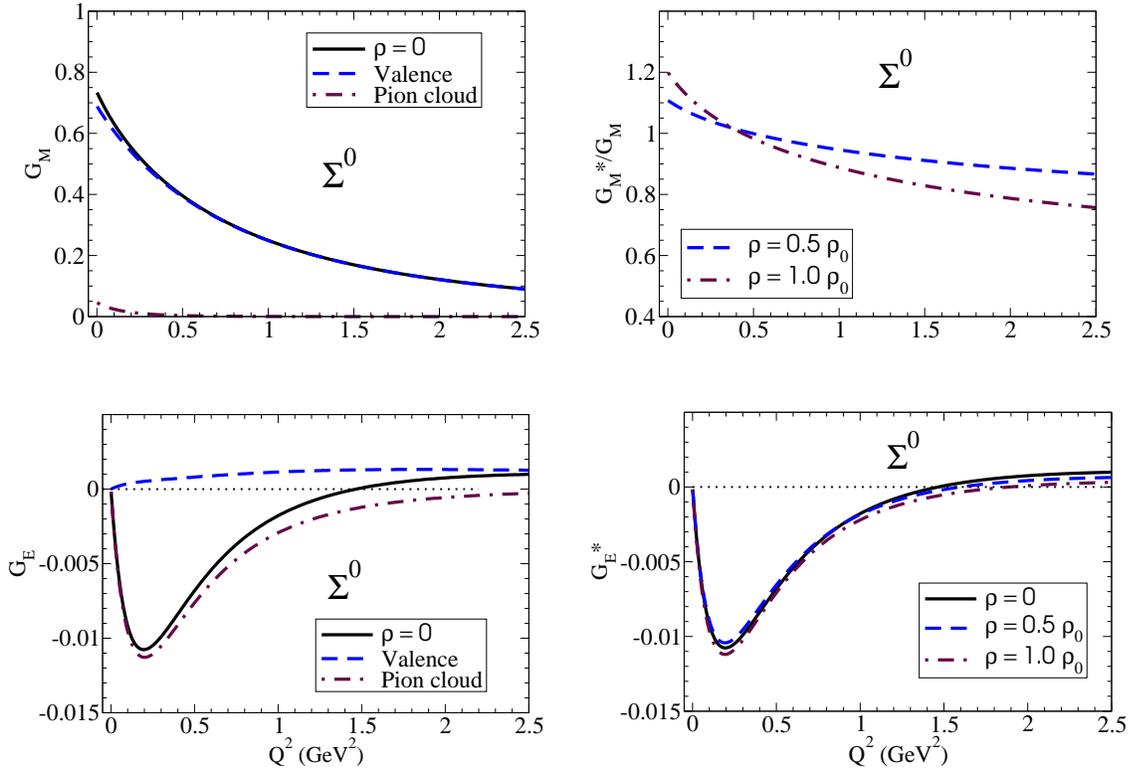

\centerline{\vspace{0.4cm}  }
\centerline{
\mbox{
\includegraphics[width=2.7in]{GMsig0.eps}   \hspace{.5cm}
\includegraphics[width=2.7in]{GMsig0R.eps}
}}
\centerline{
\vspace{0.5cm}  }
\centerline{
\mbox{
\includegraphics[width=2.8in]{GEsig0.eps}  \hspace{.4cm}
\includegraphics[width=2.8in]{GEsig0R.eps}
}}
\caption{\footnotesize{
Same as the caption of FIG.~\ref{figProton},
but for $\Sigma^0$.
}}
\label{figSigma0}
\end{figure*}

Before extending the model for the in-medium regime,
we discuss the sensitivity of the model
to the input data.
As already mentioned,
we restrict the range of variation
of $r_{E \Sigma^-}^2$ to $0.610 \pm 0.015$ fm$^2$,
instead of using the experimental result $0.61 \pm 0.15$ fm$^2$~\cite{PDG}.
The main effect of this constraint is
the reduction of the pion cloud contribution
for $r_{E\Sigma^-}^2$ to a value $\approx 0.2$ fm$^2$,
similar to that for the proton.
Also the contribution for the $\Sigma^+$ electric square radius,
(less than twice the value for $\Sigma^-$)
is similar in magnitude.
We can then conclude that the parameters
of the model are very sensitive to the
values of $r_{E\Sigma^-}^2$  and $r_{E\Sigma^+}^2$,
and the respective contributions of the pion cloud.
Thus, precise measurements
of $r_{E\Sigma^-}^2$  and $r_{E\Sigma^+}^2$ should be
very useful to improve the quality of the model,
and also to check if the assumption
$r_{E\Sigma^-}^2 \approx r_{Ep}^2$,
is justified or not.

\begin{figure*}[t]
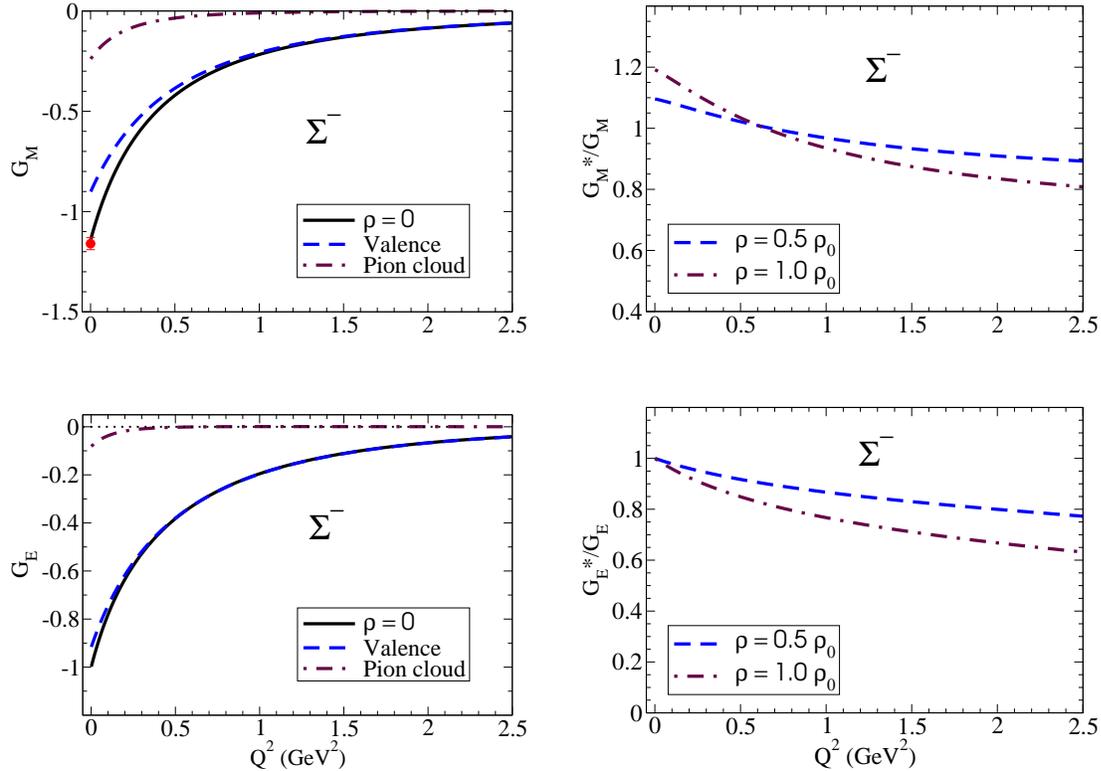

\centerline{\vspace{0.4cm}  }
\centerline{
\mbox{
\includegraphics[width=2.7in]{GMsigM.eps}   \hspace{.5cm}
\includegraphics[width=2.7in]{GMsigMR.eps}
}}
\centerline{
\vspace{0.5cm}  }
\centerline{
\mbox{
\includegraphics[width=2.7in]{GEsigM.eps}  \hspace{.5cm}
\includegraphics[width=2.7in]{GEsigMR.eps}
}}
\caption{\footnotesize{
Same as the caption of FIG.~\ref{figProton},
but for $\Sigma^-$.
The experimental magnetic moment is shown by the (red) dot.
}}
\label{figSigmaM}
\end{figure*}

A few notes about the quality of the results for
the octet baryon results are in order.
The quality of the fit is better for the systems
with a strange quark, $\Lambda,\Sigma$,
than the ones with two strange quarks, $\Xi$.
We recall that the kaon cloud contribution
neglected for the $\Xi$-baryon in the present approach,
may be important since the contribution of the pion cloud is small.
Therefore the prediction for $\Sigma$
are expected to be more reliable than that for $\Xi$.
Also, as no lattice data for $\Lambda$ and $\Sigma^0$
are used in the calibration of the model,
the results for the charge neutral particles,
$\Lambda,\Sigma^0,\Xi^0$, have to be taken with caution,
particularly for the separation of
valence and pion cloud contributions as discussed about the radii.
Nevertheless, we present the results
for the all charge neutral particles
for completeness.
Finally, we note that the predictions
for the high $Q^2$ region has also to be taken with care,
since the lattice data used in the calibration
are restricted to $Q^2 <$ 1.5 GeV$^2$.
The nucleon case is an exception
(see figure \ref{figNucleon2}),
since (physical) data are available for high $Q^2$.

\begin{figure*}[t]
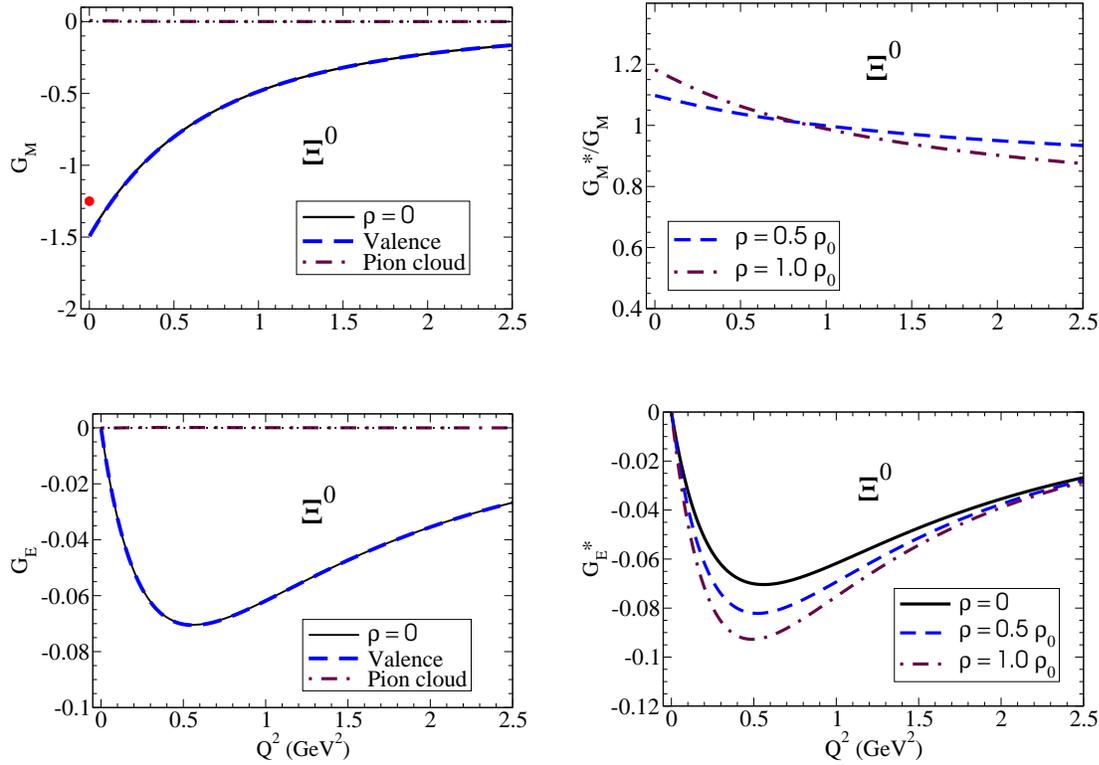

\centerline{\vspace{0.4cm}  }
\centerline{
\mbox{
\includegraphics[width=2.7in]{GMxi0.eps}   \hspace{.5cm}
\includegraphics[width=2.7in]{GMxi0R.eps}
}}
\centerline{
\vspace{0.5cm}  }
\centerline{
\mbox{
\includegraphics[width=2.7in]{GExi0.eps}  \hspace{.5cm}
\includegraphics[width=2.7in]{GExi0R.eps}
}}
\caption{\footnotesize{
Same as the caption of FIG.~\ref{figProton},
but for $\Xi^0$.
The experimental magnetic moment is shown by the (red) dot.
}}
\label{figXi0}
\end{figure*}

\subsection{Octet baryon electromagnetic form factors in medium}

We consider now the octet baryon
electromagnetic form factors in the nuclear medium.
The formalism and the parameters necessary
have already been presented in section \ref{secInMediumRegime}.
We calculate $G_E^\ast$ and $G_M^\ast$
for nuclear matter densities
$\rho=0.5\, \rho_0$ and $\rho=\rho_0$
with $\rho=0.15$ fm$^{-3}$.
The modification of the form factors in the nuclear medium
may be characterized by the in-medium modified masses given in
Table~\ref{tableMass}, and the modified coefficients
$\beta_B^\ast$ given in Table~\ref{tableCoupling}.

\begin{figure*}[t]
\centerline{\vspace{0.4cm}  }
\centerline{
\mbox{
\includegraphics[width=2.7in]{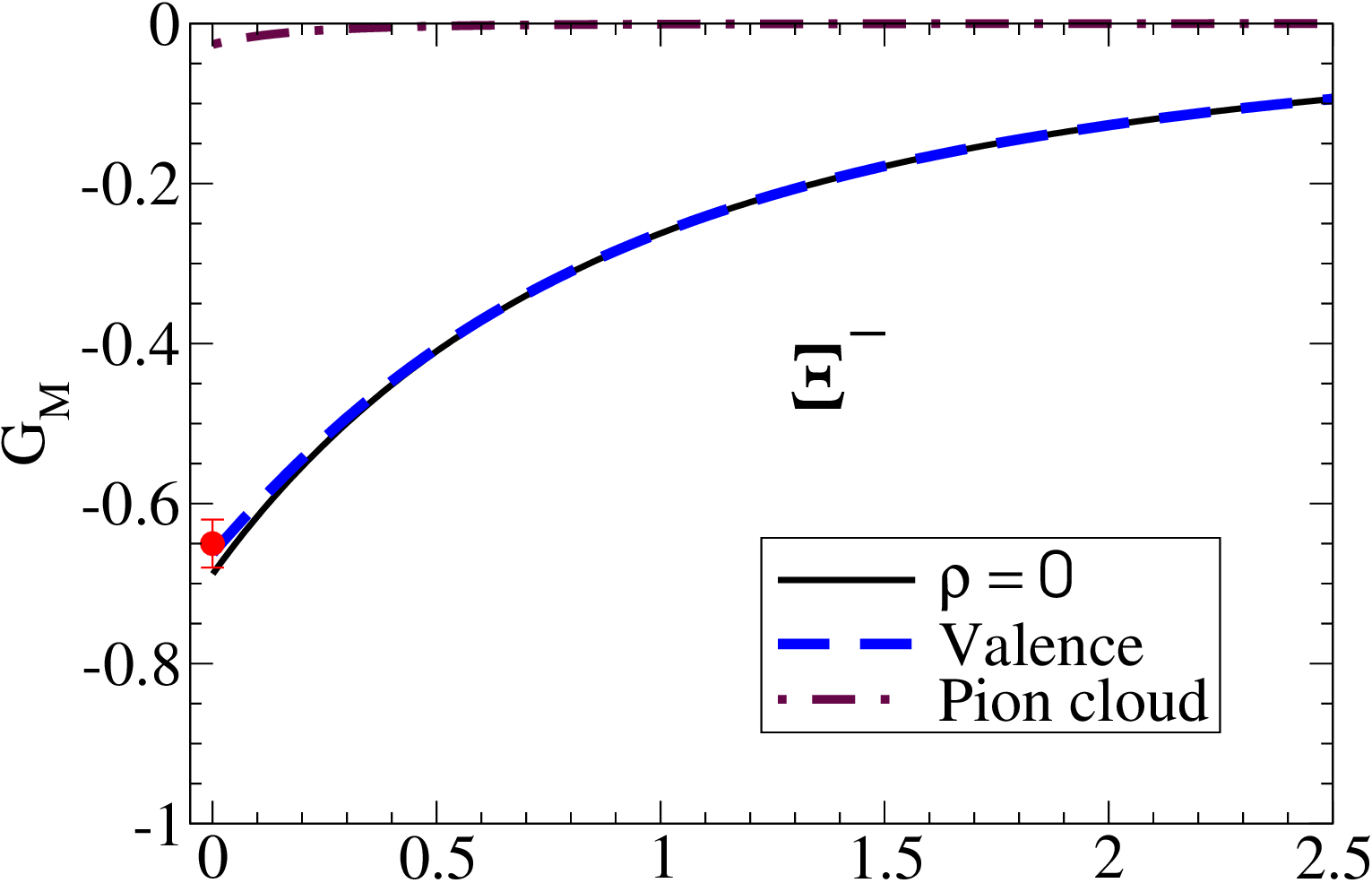}   \hspace{.5cm}
\includegraphics[width=2.7in]{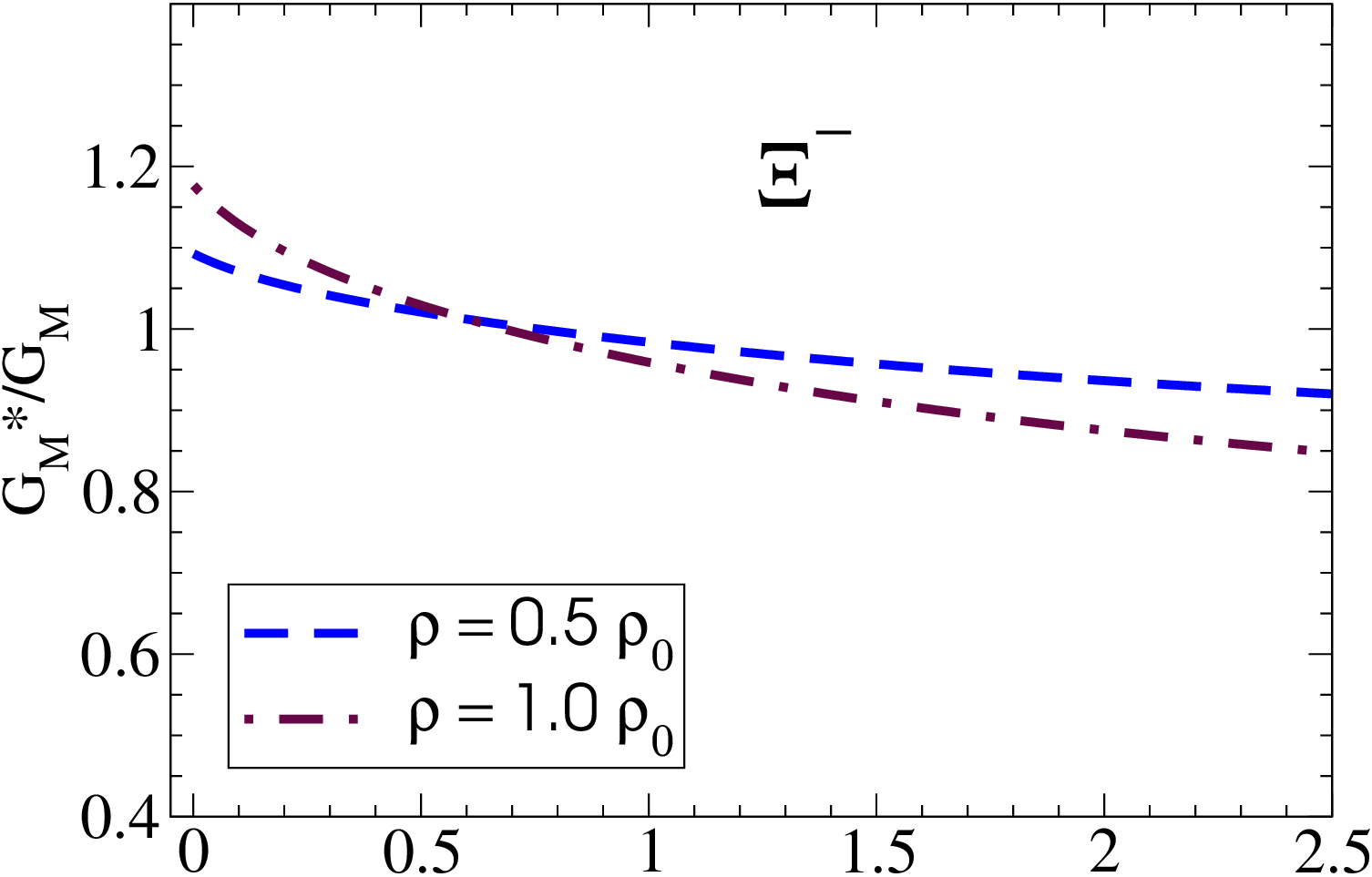}
}}
\centerline{
\vspace{0.5cm}  }
\centerline{
\mbox{
\includegraphics[width=2.7in]{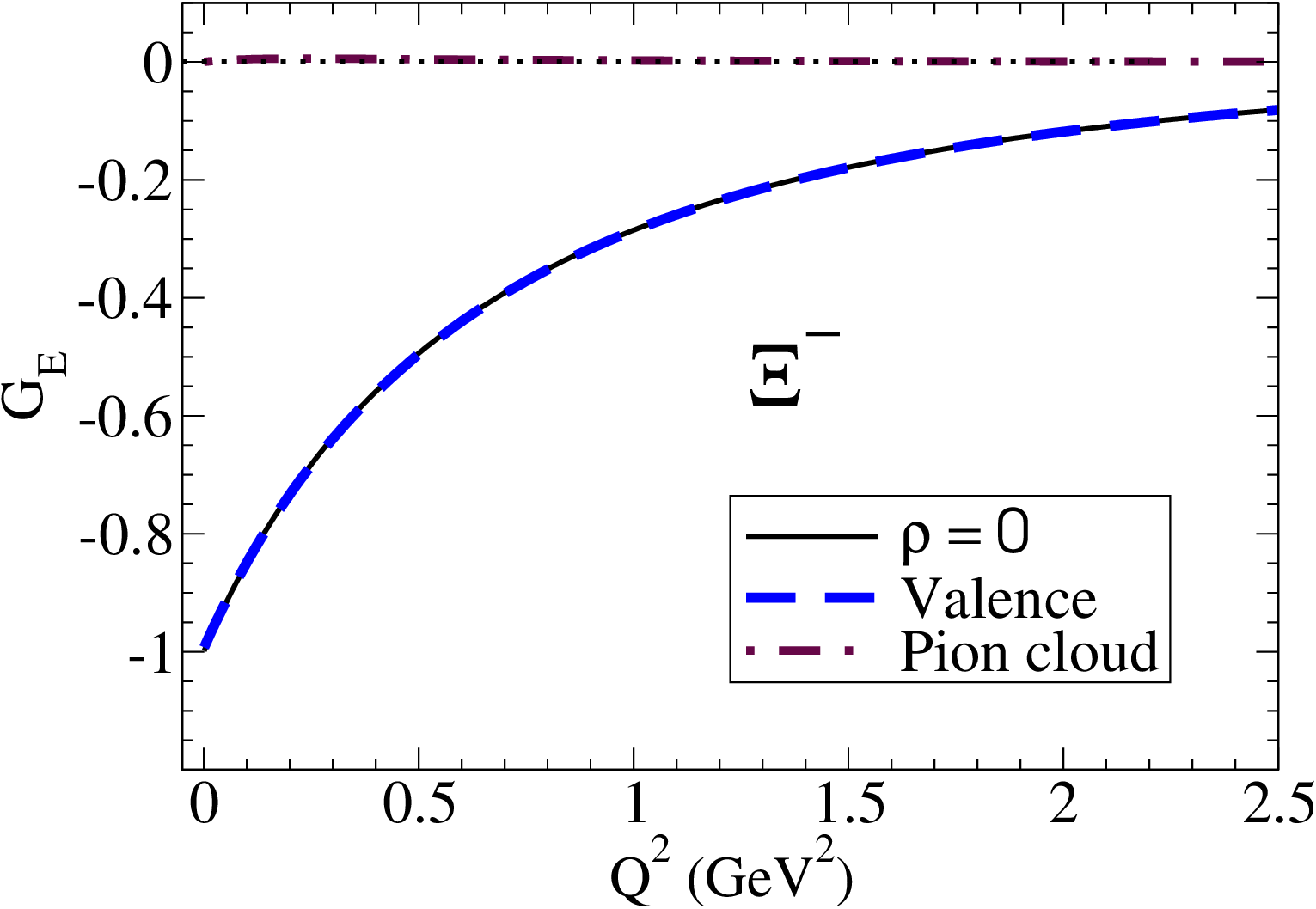}  \hspace{.5cm}
\includegraphics[width=2.7in]{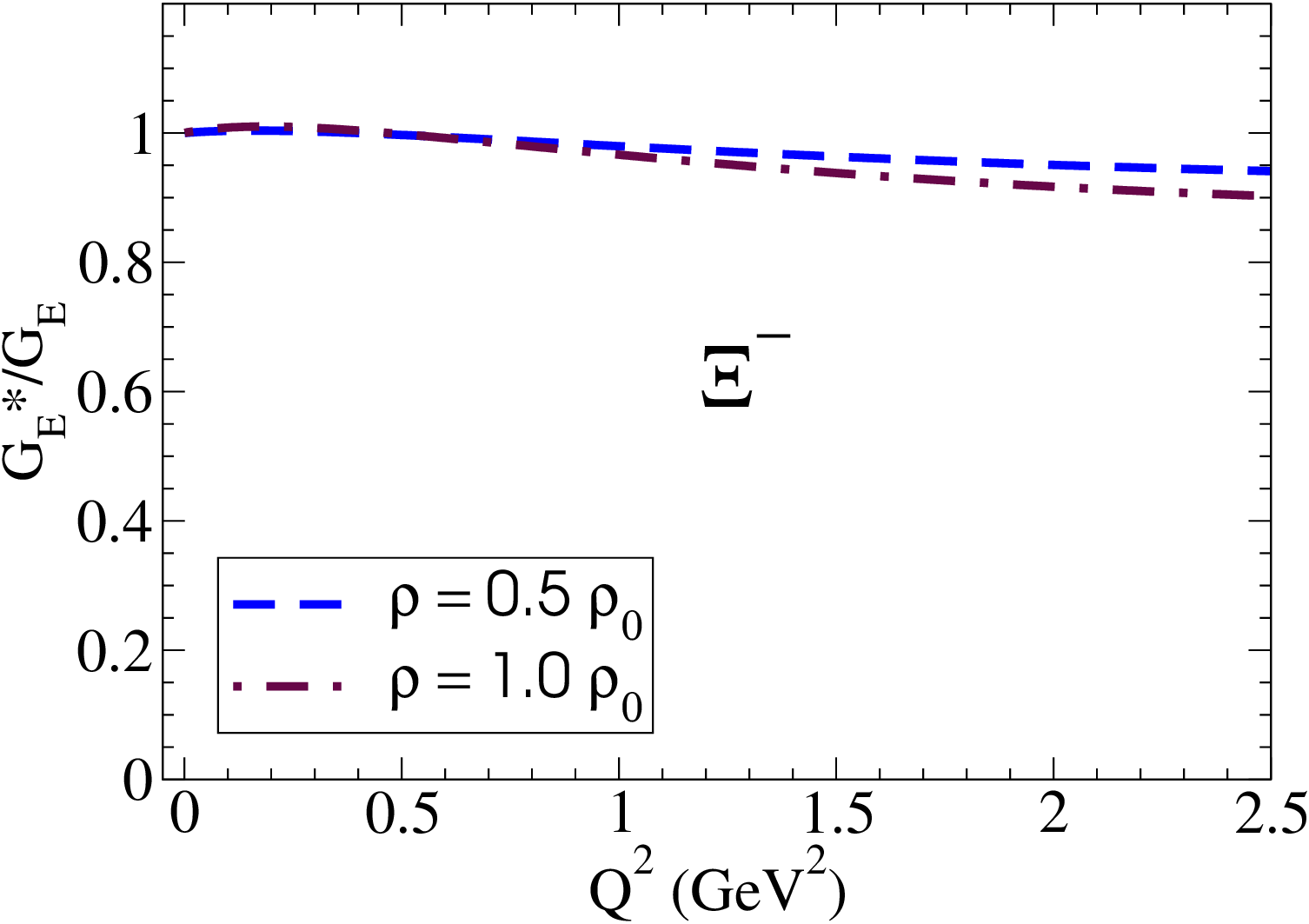}
}}
\caption{\footnotesize{
Same as the caption of FIG.~\ref{figProton},
but for $\Xi^0$.
The experimental magnetic moment is shown by the (red) dot.
}}
\label{figXiM}
\end{figure*}

The results are presented in figures \ref{figProton}-\ref{figXiM}.
On the left panel in each figure the results in vacuum
($\rho=0$) are presented, where also the valence quark contribution
(dashed line) and the pion cloud contribution (dash-dot line)
are shown. On the right panel, in-medium to vacuum ratios of
the form factors are shown. For the charge
neutral particles ($n, \Lambda,\Sigma^0,\Xi^0$),
instead, the absolute values of $G_E^\ast(Q^2)$
are shown, since $G_E(0)=G_E^\ast(0)=0$ and because the values
are in general small for finite $Q^2$ compared to $G_M^\ast(Q^2)$.
For these cases the vacuum values
are presented with the thick-solid lines.
The experimental magnetic moments in vacuum~\cite{PDG}
are also shown for the cases $B=\Lambda,\Sigma^{+,-}, \Xi^{0,-}$
with the filled circles.

From the figures it is clear that
generally the valence quark contributions are
more than 80\% of the total contribution
of each octet baryon form factor, and this
is in agreement with the results of  \cite{OctetEMFF}.
The exceptions are the electric form factors of the
charge neutral particles $n, \Lambda,\Sigma^0$ and $\Xi^0$.
To estimate the pion cloud effects
in the nuclear medium,
we note that the ratio
of the pion coupling constant in-medium to vacuum,
$g_{\pi B B'}^\ast/g_{\pi B B'}$,
given in  Table~\ref{tableCoupling} is at most 10\%.
As the factors $\beta_B$ depend quadratically
from the ratio $g_{\pi B B'}^\ast/g_{\pi B B'}$,
the effect of the variation in medium can
be up to 20\%.
Since the largest case of the pion cloud contribution
in vacuum electromagnetic form factors is about 20\%
(aside from the electric form factors of
the neutral baryons), the total modification in medium
associated with the pion cloud is at most 4\%
(20\% of 20\%).
Thus, the pion cloud effects in the nuclear medium
are essentially the same as those in vacuum,
which are shown in the left panels in
figures \ref{figProton}-\ref{figXiM}.
Thus, the significant modification
of the form factors in medium which is shown in the right
panels in figures \ref{figProton}-\ref{figXiM},
is according to the in-medium modification of
the valence quark contributions, which are
much more sensitive to the in-medium modification
than those of the pion cloud.
The strong sensitivity of the valence
quark contributions is a consequence
of the medium modification of the vector meson masses
(modification of the VMD based quark current)
and the baryon masses, which are in the radial wavefunctions.

\begin{figure*}[t]
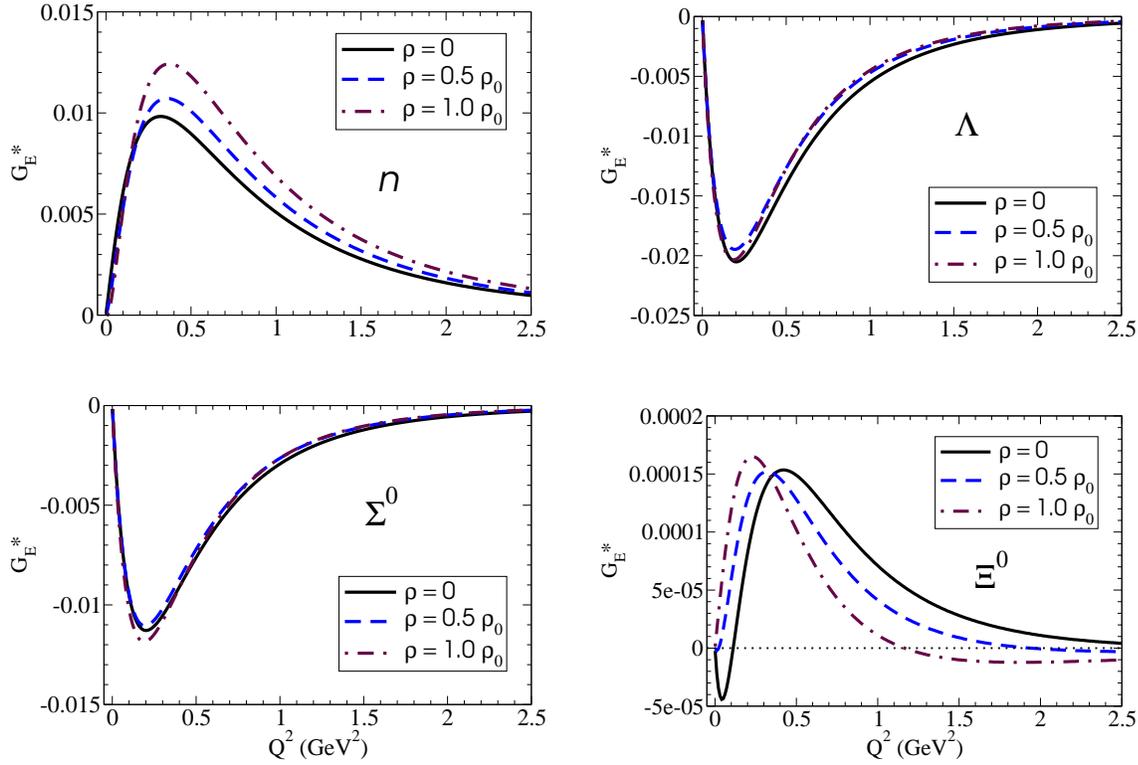

\centerline{\vspace{0.4cm}  }
\centerline{
\mbox{
\includegraphics[width=2.8in]{GEnPI.eps}   \hspace{.5cm}
\includegraphics[width=2.8in]{GElambPI.eps}
}}
\vspace{0.75cm}
\centerline{
\mbox{
\includegraphics[width=2.8in]{GEsig0PI.eps}  \hspace{.5cm}
\includegraphics[width=2.8in]{GExi0PI.eps}
}}
\caption{\footnotesize{Pion cloud contributions
for $G_E^\ast$ in nuclear matter for charge neutral particles.}}
\label{figPC}
\end{figure*}

However, the smallness of in-medium change of the pion cloud contributions
for the electromagnetic form factors (at most 4\%),
does not apply for the electric form factors of
$n,\Lambda,\Sigma^0$ and $\Xi^0$, since a priori
these electric form factors are very small
$G_E^\ast \approx Q^2$ near $Q^2 =0$.
In these cases it would be much informative
to look the direct contribution of the
pion cloud contribution in medium, not the ratio,
since it can be dominant.
In this way, we show the pion cloud contributions in vacuum and
in medium in figure \ref{figPC}.
By comparing the scales we can conclude
that the pion cloud contributions are small
for the neutron and $\Xi^0$,
but they are the leading contributions for $\Lambda$ and $\Sigma^0$.
For the neutron the contributions are about 0.01,
to be compared with the total magnitude
0.06 (see figure \ref{figNeutron}).
As for the $\Xi^0$, the pion cloud contributions are
negligible (compare the scales of $G_E^\ast$
in figure \ref{figXi0} and figure \ref{figPC}),
which is a consequence of the small coupling constant
($\beta_\Xi \approx 0.04$).
Finally for $\Lambda$ and $\Sigma^0$,
as shown in figures \ref{figLambda}
and~\ref{figSigma0}, the pion cloud
gives dominant contributions in the region $Q^2< 1$ GeV$^2$,
although the valence quarks
become dominant for higher $Q^2$.
Thus, we conclude that, although the pion cloud
gives dominant contributions for $\Lambda$ and $\Sigma^0$,
the medium modification of the form factors
according to the pion cloud is small.

We look now for the global result in medium.
For the electric form factors
we notice that the ratio $G_E^\ast/G_E$ decreases from unity
for increasing $Q^2$
(except the discussions for the charge neutral particles),
which means that the electric form factors in medium
decrease faster than those in vacuum.
The medium effect is very small for $\Xi^-$
(ratio is almost constant) reflecting
the lower sensitivity of the strange quark
to the medium modifications).
As for the magnetic form factors
they are enhanced by the medium for small $Q^2$.
The effect is larger for larger densities.
In  \cite{QMCEMFFMedium}
proton electromagnetic form factors
in medium were studied.
The results presented in  \cite{QMCEMFFMedium}
show similar trends to those observed in the present
study. Namely, the in-medium magnetic form factor
is enhanced, while that of the electric form factor
is quenched, where these results can explain
the modification of the
bound proton form factors measured at
Jefferson Lab~\cite{Strauch03}.
The modifications of the nucleon electromagnetic
form factors in medium will be discussed in next subsection.

In the literature we find only  \cite{Ryu:2008st}
studied medium modification of the
octet baryon magnetic moments aside from the nucleon,
using a different kind of quark-meson coupling model
from the present one,
``QMC'' and modified quark-meson coupling model (MQMC).
In their treatment, the pion or meson cloud is not
included in the electromagnetic currents.
(We discuss the nucleon case in next section.)
They compared with ``QMC'' and MQMC results for the
magnetic moments in nuclear medium.
In their ``QMC'', the in-medium magnetic moments
for the $\Lambda$ and $\Xi^-$ decrease compared
to those in vacuum, although other octet magnetic moments are
all increased in medium.
This feature is different from the present approach, where
all the octet baryon magnetic moments ($G^\ast_M(Q^2)$ at $Q^2=0$)
are enhanced in medium.
In their MQMC, on the other hand, all the octet magnetic
moments are enhanced, and this feature is the same as
that of the present approach.

\subsubsection{Nucleon electromagnetic form factors in-medium.}

We study now in detail the medium modification
of the nucleon electromagnetic form factors.
A very interesting quantity is the ratio
of the electric to the magnetic form factors,
\ba
R_N^\ast = \frac{G_{EN}^\ast}{G_{MN}^\ast},
\ea
which can be also calculated in vacuum (denoted by $R_N$).
The study of this ratio for
the proton in vacuum serves as a fundamental
quantity in the present understanding of the proton
structure, as was measured at Jefferson Lab,
that $R_p$ significantly
deviates from a constant~\cite{JlabR1,Puckett10}.
Similar studies were also made for the
neutron~\cite{Madey03,Riordan10,Plaster06}.

\begin{figure*}[t]
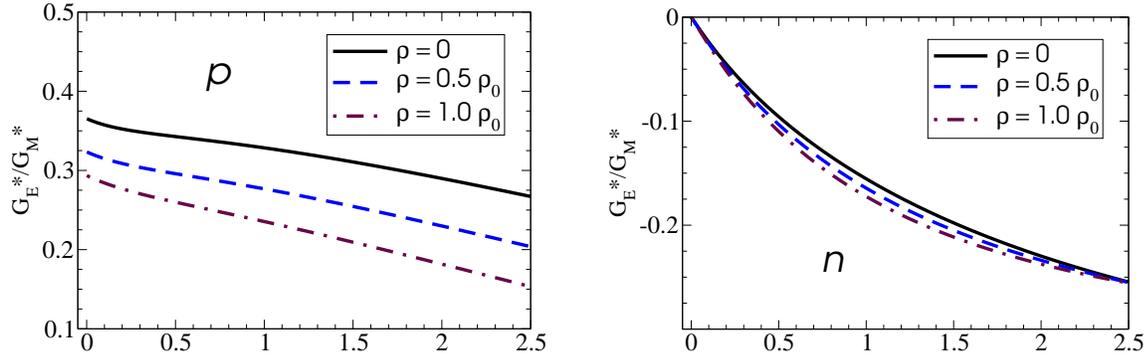

\centerline{\vspace{0.4cm}  }
\centerline{
\mbox{
\includegraphics[width=2.8in]{ProtonGEGM.eps} \hspace{.6cm}
\includegraphics[width=2.8in]{NeutronGEGM.eps} }}
\caption{\footnotesize{
Ratios $G_E^\ast/G_M^\ast$ in nuclear matter for the proton
(left) and neutron (right).}}
\label{figNucleonSR}
\end{figure*}

In figure \ref{figNucleonSR} we present our
predictions for the both,
proton and neutron ratios in medium, for
nuclear densities $\rho=0.5 \, \rho_0$
and $\rho=1.0\, \rho_0$.
Note that $\rho=0$ (vacuum) case was
already compared with the experimental data in
the present model since it fits both
$G_E$ and $G_M$ (see figure \ref{figNucleon}).
Experimentally, direct access for the ratio $R^\ast_N$
in-medium seems to be impossible at present.
However, we can get indirect information for
the in-medium ratios using the results of proton (nucleon)
recoil polarization experiments in nuclei by the measurement
of the polarization-transfer super-ratio for the nucleon,
\ba
{\cal R} = \frac{G_E^\ast/G_M^\ast}{G_E/G_M}.
\label{eqSR}
\ea
The experiments were performed
for the proton using $^1$H, $^4$He, and
$^{16}$O targets~\cite{Dieterich01,Strauch03,Paolone10},
and also planed for the neutron case~\cite{Cloet09,npolarization}.

\begin{figure*}[t]
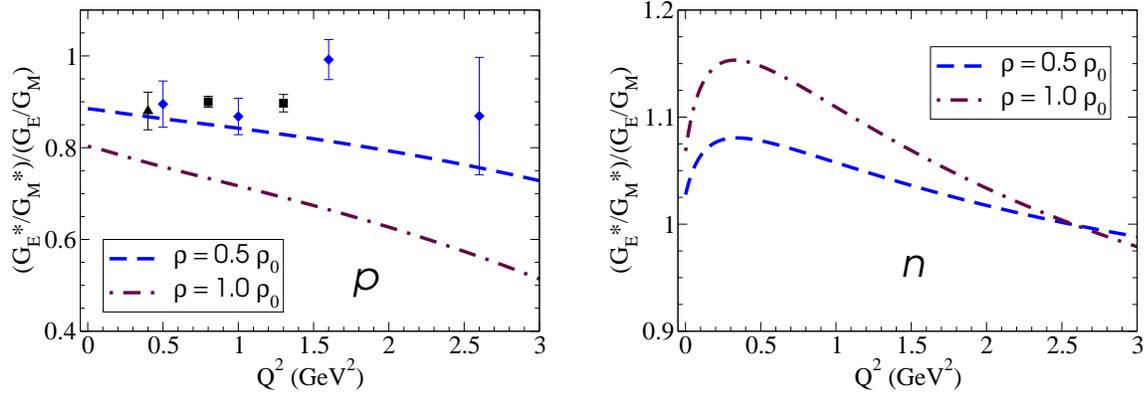

\centerline{\vspace{0.5cm}  }
\centerline{
\mbox{
\includegraphics[width=2.8in]{ProtonDRatio3.eps} \hspace{.6cm}
\includegraphics[width=2.8in]{NeutronDRatio2.eps} }}
\caption{\footnotesize{
Super-ratios in nuclear matter calculated for the proton (left)
and neutron (right).
Data for proton are from  \cite{Dieterich01,Strauch03,Paolone10}.
}}
\label{figProtonDR}
\end{figure*}

The results of our predictions for ${\cal R}$
for the proton and the neutron are
presented in figure \ref{figProtonDR}.
Some other calculations for the proton case can be found
in  \cite{QMCEMFFMedium,Miller,MEC,Udias99,Laget94,Caballero97,Udias00,Lava04},
and summarized in  \cite{JLabbook,npolarization}.
For a detailed discussions about the super-ratio
of the proton and the neutron, see \cite{JLabbook,Cloet09,Chung91}.

First, we discuss the proton super-ratio shown
in the left panel in figure \ref{figProtonDR} with
the data with the $^4$He target from
 \cite{Dieterich01,Strauch03,Paolone10}.
Our results for $\rho=0.5\, \rho_0$ ($\rho_0=0.15$ fm$^{-3}$)
reproduce the data. While the average nuclear density
of $^4$He is expected to be slightly higher
($0.74 \,\rho_0$ in a QMC estimate)
the effect of absorption means that the active nucleon
tents to be in the nuclear surface.
Thus, the present model predicts the observed
trend of the reduction for the super-ratio.

Next, turning the discussion to the neutron case,
our results predict an {\it enhancement} of the
super-ratio, contrary to the case of the proton.
The enhancement was also predicted by
Nambu-Jona-Lasinio (NJL) model and relativistic
light front constituent quark model (LFCQ)~\cite{Cloet09}.
However, the difference is the $Q^2$ dependence.
In the case of the NJL model, the super-ratio monotonically decreases
with increasing $Q^2$, while it stays almost constant in LFCQ.
The present result shows appreciable $Q^2$ dependence,
namely the ratio  increases up to around $Q^2=0.3$ GeV$^2$,
and gradually decreases with increasing $Q^2$.
Thus, our model predicts that the modification
can be maximally observed around $Q^2=0.3$ GeV$^2$.
This point may be taken into consideration
in the experiments planned~\cite{npolarization}. 

\section{Summary and conclusions}
\label{secDiscussions}

In this work we have presented a model for the octet
baryon electromagnetic form factors in vacuum
and in the nuclear medium.
The model is based on a constituent quark formalism
but includes also a phenomenological parametrization
for the pion cloud motivated by $\chi$PT.
The octet baryon and relevant meson properties in
the nuclear medium (masses and coupling constants)
are determined using the QMC model.
The effects of final state interactions and meson exchange current
are not explicitly included.

The model is calibrated by the octet baryon
lattice data (electromagnetic form factors),
as well as physical data, such as
the nucleon electromagnetic form factors,
octet baryon magnetic moments and the
available octet baryon radii data.
Lattice data give
stringent constraints
for the valence quark structure of the octet baryons.
The remaining data combined with
the estimate of the valence quark core
contributions, constrain the parametrization
of the pion cloud.

The fit is very sensitive to the
neutron lattice data for the electric
form factor and also to the value of the $\Sigma^-$
electric radius.
The neutron data, lattice and physical, are well
described by a valence quark model
with isospin breaking.
An accurate description of the data, including
the octet radii (nucleon and $\Sigma^-$)
is also obtained with a small value
for electric square radii for $\Sigma^-$
($\approx 0.6$ fm$^2$)
which is a consequence of a small pion cloud
contribution for the electric radius
(same order as that for the proton).

Future improvements are possible.
Accurate lattice data, particularly for
the neutron electric form factor ($G_{En}$)
can clarify the role
of the valence quark contributions,
and eventually demand a refit of
the quark current used in this work.
The available lattice data support a
quark electromagnetic current with
an isospin breaking,
but new data with a smaller effect for
$G_{En}$ may require a model with almost no isospin breaking.
The model can also be improved by including a
pion cloud parametrization derived from
first principle QCD,
utilizing the results of lattice simulations
(see Ref.~\cite{Wang09}).

The explicit inclusion of meson exchange current corrections
in the electromagnetic form factors in a consistent manner,
may be appreciable at high momentum transfer,
especially 
if the absolute values of form factors became
very small, 
or for the electric form factors of neutral baryons.

We predict that all the octet baryon magnetic form factors
in medium will be enhanced in the small $Q^2$ region,
and the magnetic moments of the octet baryons in medium
are enhanced.
The model predicts that $Q^2$ dependence of the octet baryon electric
form factors in medium decreases faster than those in vacuum.
Furthermore, the present model predicts also a decrease of the
super-ratio for the proton in the nuclear medium, which
is in agreement with the observed results.
On the other hand, for the neutron super-ratio in the nuclear medium, the present
model predicts the {\it enhancement}, which has its maximum
around $Q^2=0.3$ GeV$^2$,
and this may be useful information for the planned experiments.

\vspace{0.2cm}
\noindent
{\bf Acknowledgments:}

We thank K.~Saito for useful discussions
and warm hospitality (K.T.) at Tokyo University of Science,
Noda, Japan, where part of the work was undertaken.
G.R.~would like to acknowledge
CSSM at the University of the Adelaide for making
it possible for him to visit and stay.
This work was supported in part by the European Union
(HadronPhysics2 project ``Study of strongly interacting matter''),
by the University of Adelaide and
by the Australian Research Council through grant No.~FL0992247 (AWT).
G.R.\ was supported by the Funda\c{c}\~ao para
a Ci\^encia e a Tecnologia under the Grant
No.~SFRH/BPD/26886/2006.
K.T.\ would like to acknowledge the International Institute of Physics,
Federal University of Rio Grande do Norte, Natal, Brazil,
for a visiting professorship during which part of this work
was carried out.

\end{document}